\documentclass[preprint,12pt]{elsarticle}

\usepackage{graphicx}
\usepackage{multirow}
\usepackage{amssymb}
\usepackage{amsmath}
\usepackage{txfonts}
\usepackage{appendix}
\usepackage{pdflscape}
\usepackage{longtable}

\usepackage{url}
\usepackage[reviewcopy]{adndt}

\biboptions{comma,sort&compress}

\journal{Atomic Data and Nuclear Data Tables}

\begin{document}

\begin{frontmatter}


\title{A Critical Compilation of Experimental Data on the 1s2$l$2$l^{\prime}$ Core-Excited States of Li-like Ions from Carbon to Uranium}



\author[NIST1]{V.I. Azarov\corref{cor1}}
\author[NIST]{A. Kramida\corref{cor2}}
\author[NIST]{Yu. Ralchenko\corref{cor3}}

\cortext[cor1]{Permanent address: Department of Research and Development, Parametric Technology Corporation, 121 Seaport Boulevard, Boston, 02210, USA. E-mail: vlad\_azarov@yahoo.com}
\cortext[cor2]{Corresponding author. E-mail: alexander.kramida@nist.gov}
\cortext[cor3]{E-mail: yuri.ralchenko@nist.gov}
\address[NIST1]{Associate, National Institute of Standards and Technology, Gaithersburg, MD 20899, USA}
\address[NIST]{National Institute of Standards and Technology, Gaithersburg, MD 20899, USA}

\begin{abstract}
Recent advance in calculations of energy levels of the 1s2$l$2$l^{\prime}$ core-excited states for ions along the Li isoelectronic sequence from carbon to uranium suggested that theoretical predictions for the 1s$^22l$-1s2$l$2$l^{\prime}$ transitions are significantly more precise than most of the experimental results available today and can thus be used for calibrating measured X-ray spectra. This suggestion is verified in the present work by comparison with benchmark experimental results. 
We present a critical compilation of all available experimental data on the energy levels of the 1s2$l$2$l^{\prime}$ core-excited states of Li-like ions. The compiled experimental data include uncertainty estimates, which allowed proper weighted averaging and comparison with theoretical calculations. This comparison confirmed the supremacy of the most advanced theoretical calculations of the 1s2$l$2$l^{\prime}$ levels over the best experimental results available today.
\end{abstract}


\end{frontmatter}




\newpage

\tableofcontents
\listofDtables
\listofDfigures
\vskip5pc


\section{Introduction}
\label{S:1}

Dielectronic satellite spectra of highly charged ions are an important source of spectroscopic data widely used for analyzing astrophysical plasmas  as well as spectra from electron beam ion traps, tokamaks and laser-produced plasmas (see, e.g., \cite{WOS:000233347800055,WOS:000460814000017,WOS:000226876100019,WOS:000481918000005}). Such spectra provide crucial information on the electron temperature and density, the ionization state distribution, and other characteristics of hot plasmas. An analysis of such spectra requires detailed knowledge of energy levels of various excited states of highly charged ions. High-quality energies (theoretical or experimental) are critical for a proper fit of spectral line profiles and thus for a better plasma diagnostic.

A significant progress in theory of highly charged ions has been achieved in the last years. For Li-like ions, new advanced theoretical studies provided energies with estimations of uncertainties due to effects that are not included in previous calculations \cite{Y17,Y18}. For the core-excited states of these ions, these theoretical predictions were shown to be significantly (by an order of magnitude) more precise than the best experimental energies available today. This conclusion was based on a comparison of theoretical values with benchmark experimental results for several atomic numbers $Z$ in the region from $Z = 6$ to $Z = 80$ and for several different experiments. In the study by Yerokhin et al. \cite{Y17,Y18}, energy levels and fine-structure intervals of the 1s2$l$2$l^{\prime}$ core-excited states were calculated for ions along the Li isoelectronic sequence from carbon to uranium. All theoretical energies were supplied with uncertainty estimates. 

The goal of the present work is to extend a collection of benchmark experimental results used for the verification of the theoretical calculations in Refs. \cite{Y17,Y18} to all available experimental data on the 1s2$l$2$l^{\prime}$ core-excited states for ions along the Li isoelectronic sequence, and to compare these data with the theoretical results obtained in Refs. \cite{Y17,Y18}. Validation of both the calculated energies and their uncertainties is our task. If the correctness of the statement about precision of the theoretical predictions is confirmed on all available experimental data, high-precision theoretical energies of core-excited states of Li-like ions published in Refs. \cite{Y17,Y18} may become the preferable source of spectroscopic data for modeling plasma spectra and may be used for calibration of experimental X-ray spectra of ions with a larger number of electrons. 

In this work, when discussing the satellite structures near the resonance lines of He-like ions, we use the designations of Gabriel \cite{Gabriel_1972} (a, b, c, etc.) for transitions forming these structures. Translation of these designations to standard spectroscopic notation of transitions will be discussed in the following Section. 

\section{Coupling schemes}
\label{section:coupling}
When reading the literature on the core-excited levels of Li-like ions, one faces a difficulty with heterogeneous spectroscopic notation. In the early papers \cite{Gabriel_1969,Gabriel_1972} dealing with moderately charged ions, a peculiar kind of $LS$-coupling scheme was used to designate the 1s2s$^2$, 1s2p$^2$, and 1s2s2p energy levels, which we will call Gabriel's scheme. For the 1s2p$^2$ configuration having two equivalent p-electrons it is impossible to distinguish between them, so the two p-electrons are combined first to produce an intermediate singlet ($^1$S or $^1$D) or triplet ($^3$P) term, which is then combined with the $^2$S term of the 1s shell to produce the final $LS$ term. For the odd 1s2s2p configuration, Gabriel's coupling scheme first combines the 1s and 2p electron to produce an intermediate $^1$P$^{\circ}$ or $^3$P$^{\circ}$, which then are combined with the remaining 2s electron to obtain the final doublet and quartet $LS$-terms. In many subsequent studies (see, e.g., Vainshtein and Safronova \cite{Vainshtein_1978}), a `reordered' $LS$-coupling was used, in which the two $n = 2$ electrons are combined first, and then the intermediate singlet and quartet terms are combined with the 1s-electron. Other researchers (see, e.g., Chen \cite{Chen_1986}) used a `sequential' $LS$-coupling, in which the electronic subshells are combined in the order of their binding energy, i.e., first 1s is combined with 2s, and then the 2p subshell is added. We put the word `sequential' in quotes, because, as mentioned above, the 1s2p$^2$ configuration must be an exception: the two 2p-electrons are indistinguishable and thus must be combined first, same as in Gabriel's scheme. Further advancement of both theory and experiments into the region of very high nuclear charges $Z$ led to realization that at high $Z$ the coupling changes to $jj$, which also has many different versions. Most relativistic atomic-structure codes have only a `sequential' $jj$-coupling option (with the same exception for the 1s2p$^2$ configuration). However, Cowan's codes \cite{Cowan_1981,Kramida_Cowan}, while making all calculations in $LS$-coupling, have an option to transform the output to $jj$-coupling, as well as several other coupling schemes, with a given order of summation of subshells. For subshells with equivalent electrons, such as 2p$^2$, these codes can only compute the $LS$-coupling level composition. 

\begin{figure}[ht!]
 \centering
 \includegraphics[width=.49\linewidth]{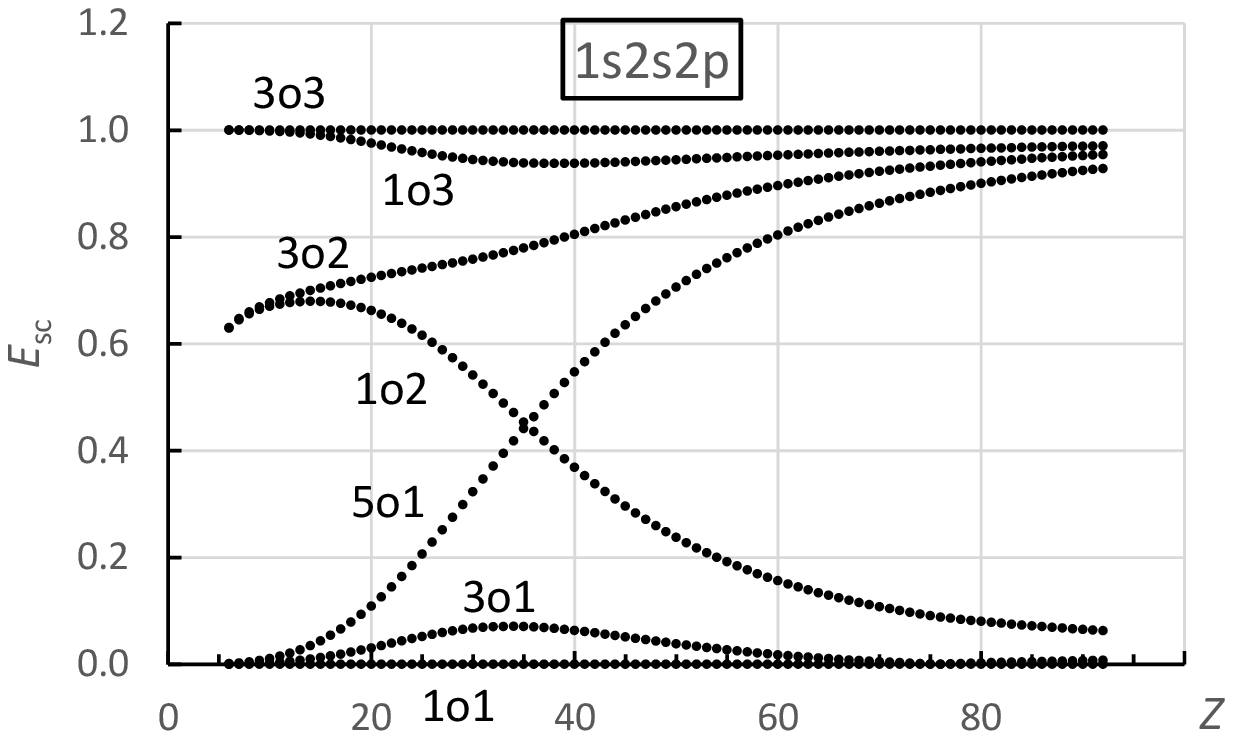}
 \includegraphics[width=.49\linewidth]{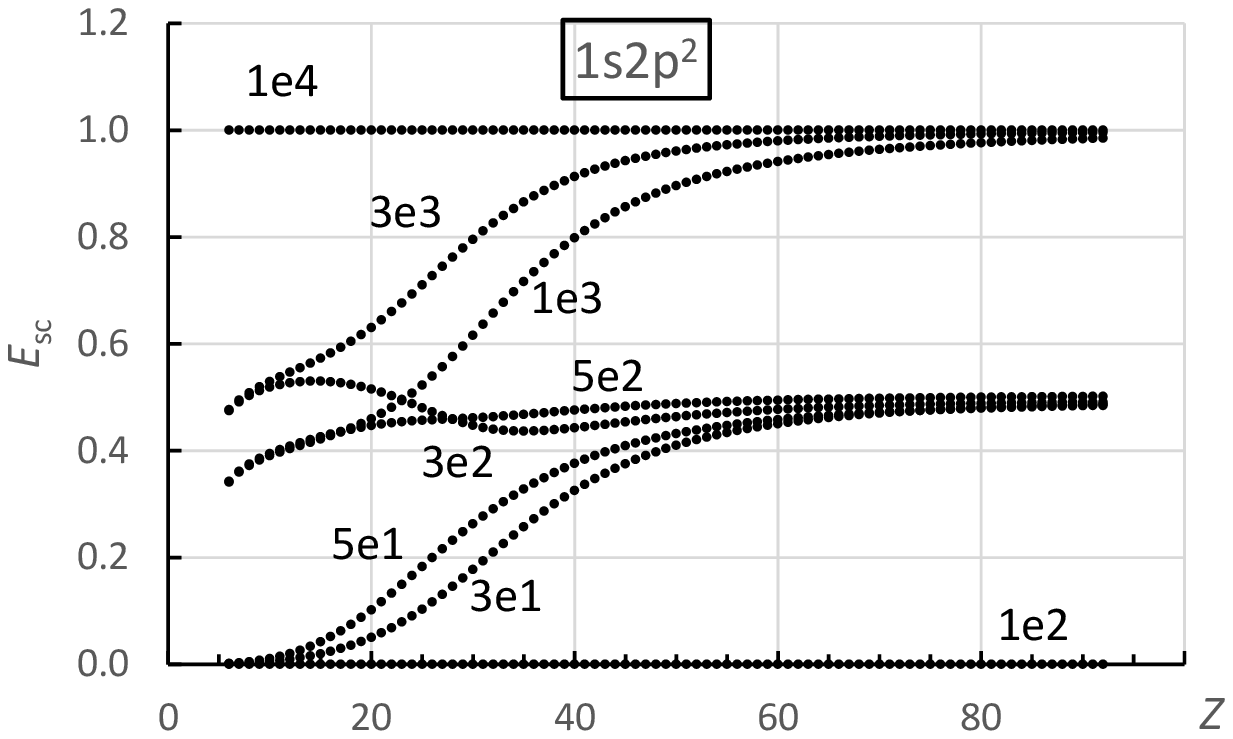} 
 \caption{Energy structure of the 1s2s2p (left) and 1s2p$^2$ (right) configurations of Li-like ions as a function of nuclear charge $Z$. $E_{\text{sc}}$ is the energy separation from the lowest level of each configuration divided by the energy spread of that configuration in each ion. For explanation of level labels, see text.}
 \label{fig:Esc}
\end{figure}

Yerokhin et al. \cite{Y17,Y18} did not provide percentage composition of their calculated energy levels. To designate the levels, for all $Z$ they used the same `sequential' $LS$-coupling labels as in C~IV ($Z = 6$). In Section 4 of Ref. \cite{Y18} the authors mentioned in passing that at high $Z$ the levels structure approaches the $jj$-coupling limit. However, the choice of a physically appropriate coupling scheme for each $Z$ was, to our knowledge, never discussed in the literature. To illustrate the problem, we depicted their scaled energies of the 1s2s2p and 1s2p$^2$ configurations in Fig. \ref{fig:Esc}. For the scaling, the energies counted from the lowest level of each configuration were divided by the energy spread of that configuration in each ion. The figure clearly shows that, indeed, the coupling conditions change drastically from the low-$Z$ to high-$Z$ end of the sequence. One can hardly expect the $LS$-coupling labels given by Yerokhin and Surzhykov \cite{Y18} to have much physical meaning for Li-like uranium. The level labels used in Fig. \ref{fig:Esc} are explained below.


The smoothness of all energy-level dependencies on $Z$ seen in Fig. \ref{fig:Esc} suggests that, in the context of isoelectronic comparisons, the energy is a crucial part of the level identity. The dominant $LS$- or $jj$-coupling character of a level may change with $Z$, but we found that the energy order of levels within each $J^{\pi}$ set (where $J$ is the total angular momentum quantum number and $\pi$ is the parity) is preserved for the entire isoelectronic sequence. Therefore, we chose to designate the levels with labels constructed from three symbols: an integer number equal to $2J$ followed by a parity symbol (`e' for even and `o' for odd) followed by sequential number of the level in the order of increasing energies in each $J^{\pi}$ set. The numbering starts with zero in the sets that include 1s$^2$2$l$ levels and from 1 otherwise, so that the first core-excited level in each set has the number 1 in its label. These labels are used in Fig. \ref{fig:Esc} and in all other figures and tables, as well as the main text.

To determine the best coupling scheme describing the level structure for each ion, we calculated percentage compositions of levels in the `sequential' and `reordered' $LS$- and $jj$-coupling schemes for $Z = 6$, 18, 26, 34, 42, 54, 68, 80, and 92. The `sequential' $jj$-coupling calculations were made with the Flexible Atomic Code (FAC) code \cite{Gu_2008}, while the `reordered' $jj$-coupling, as well as 'sequential' and `reordered' $LS$-coupling calculations were made with Cowan's codes \cite{Cowan_1981,Kramida_Cowan}. In the Cowan-code calculations, only the 1s$^2$2s, 1s$^2$2p, 1s2s$^2$, 1s2s2p, and 1s2p$^2$ configurations were included, and the Slater parameters were varied in the least-squares fitting procedure to fit the level values given by Yerokhin et al. \cite{Y17,Y18}. The FAC calculations additionally included the 1s$^2nl$ ($n = 3$, 4, 5, $l \leq n-1$), 1s(2s+2p)$nl$ ($n=3$, 4, $l \leq n-1$), and 1s(3s+3p+3d)$^2$ configurations (38 configurations in total). The only output in common to the two sets of calculations was the percentage composition in the `sequential' $jj$-coupling for the 1s2s2p configuration. For this configuration, the two sets of our results for percentage composition were found to agree within 2.4 \% on average.

\begin{figure}[ht!]
 \centering
 \includegraphics[width=.49\linewidth]{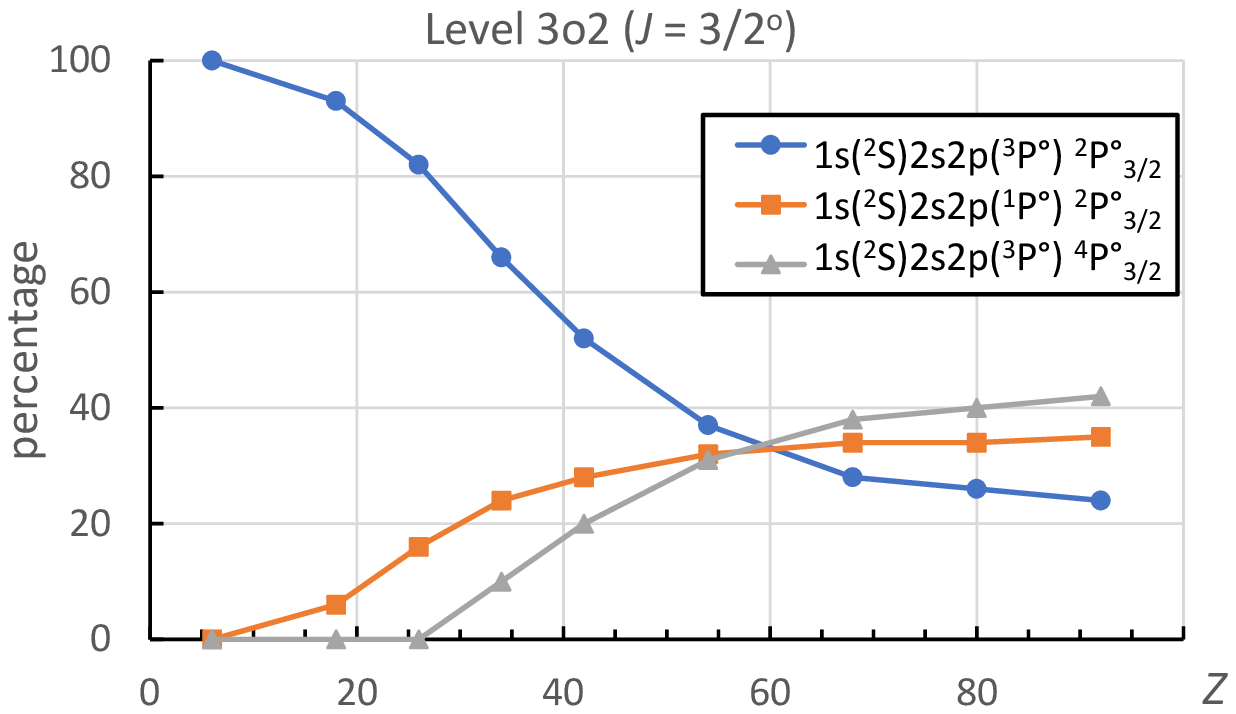}
 \includegraphics[width=.49\linewidth]{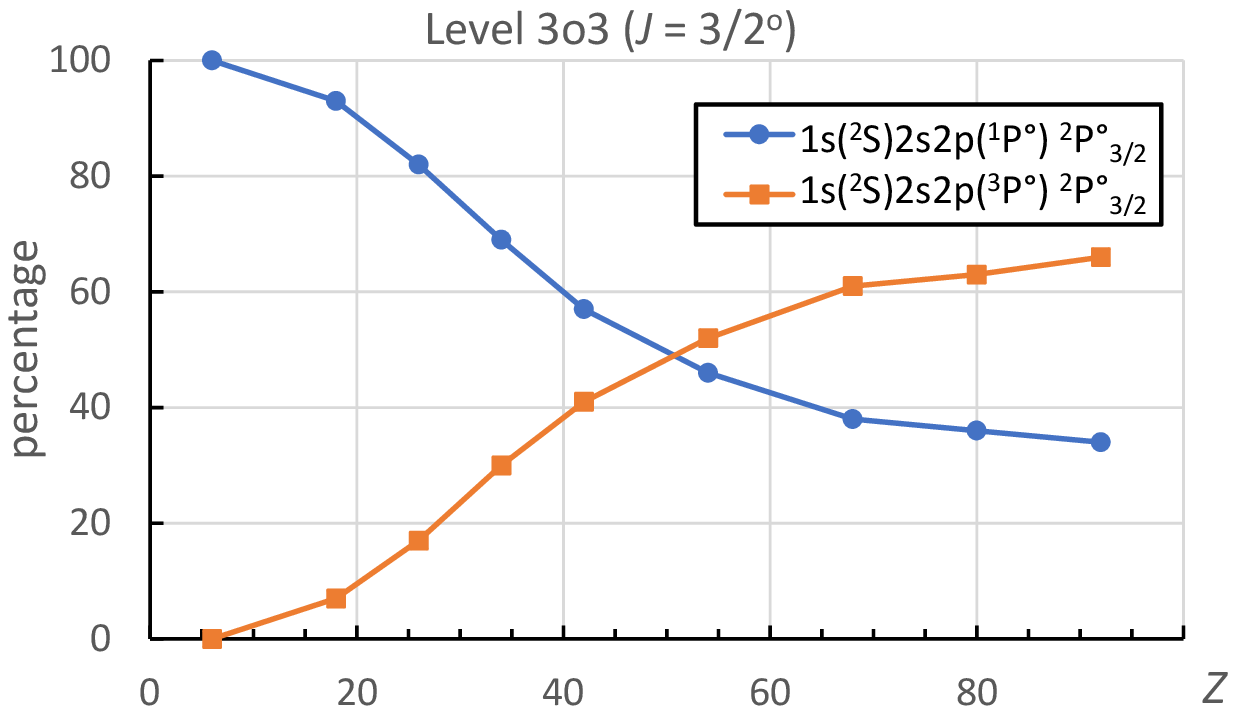}
 \includegraphics[width=.49\linewidth]{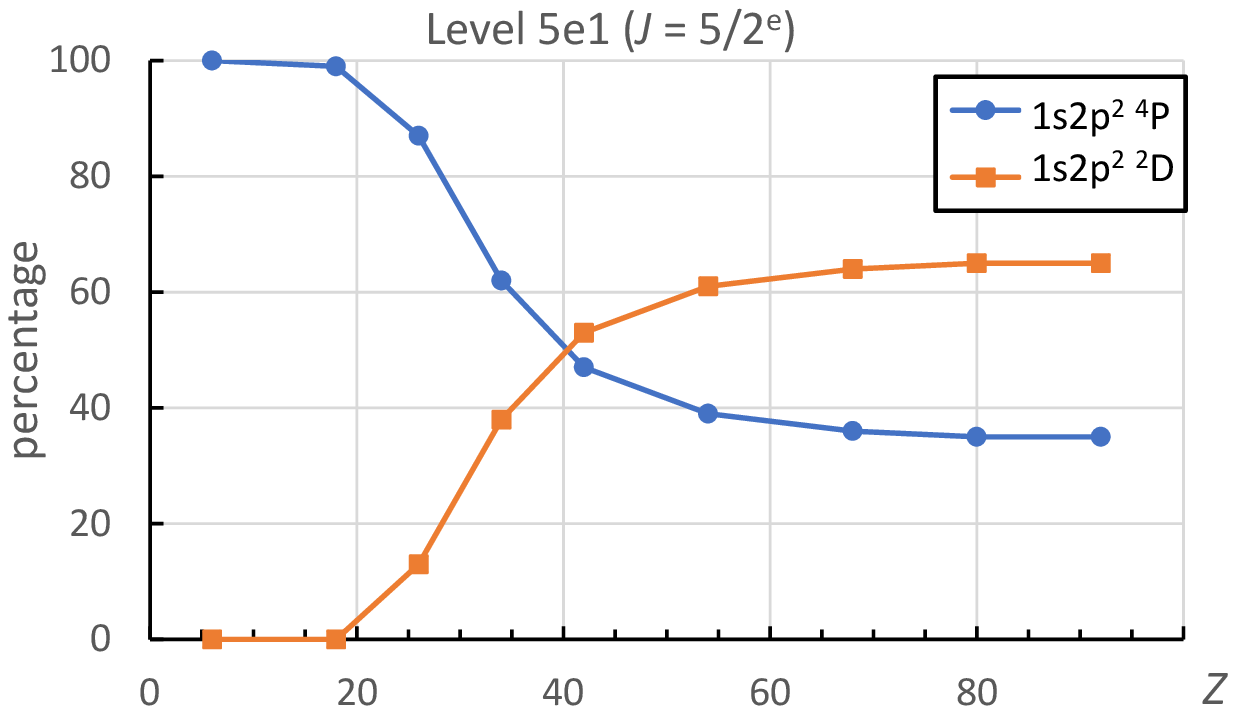}
 \caption{Percentages of the eigenvector components in the composition of two 1s2s2p $J=3/2$ levels (top) and one 1s2p$^2$ $J=5/2$ level (bottom) of Li-like ions as a function of nuclear charge $Z$. The `reordered' $LS$ coupling was used to construct the basis sets (see text).}
 \label{fig:pcnt_LS_r}
\end{figure}

The change in composition of the 1s2s2p and 1s2p$^2$ levels along the sequence in the `reordered' $LS$-coupling scheme is illustrated in Fig. \ref{fig:pcnt_LS_r}. This figure shows only the levels with the most drastic changes. For even parity, there are only two levels with $J=5/2$, so only one plot is shown. The plot for the other $J=5/2$ level, 5e2, looks exactly the same with the labels interchanged. It is these two levels, 5e1 and 5e2, that set the upper limit ($Z \leq 38$) for applicability of $LS$ coupling labels to the $n=2$ core-excited levels in Li-like ions. In odd parity, switching of the leading $LS$ components of the 3o2 and 3o3 levels happens at much greater $Z$, near $Z = 60$.

\begin{figure}[ht!]
 \centering
 \includegraphics[width=.49\linewidth]{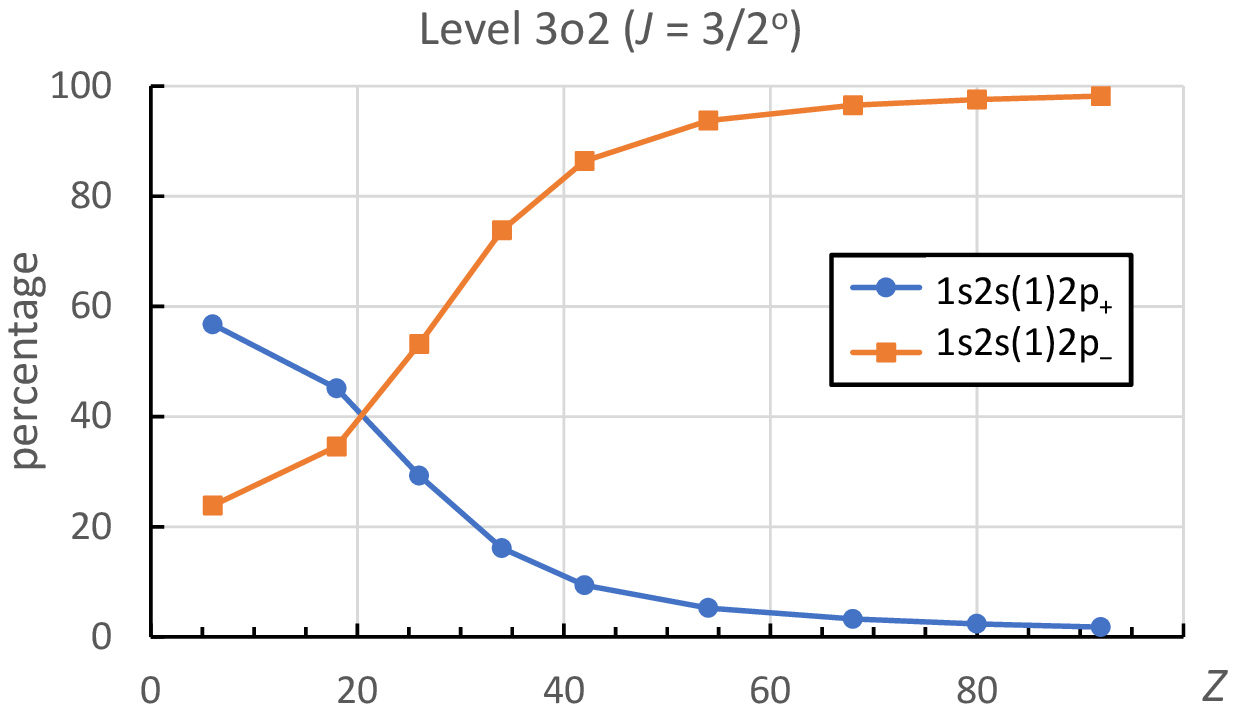}
 \includegraphics[width=.49\linewidth]{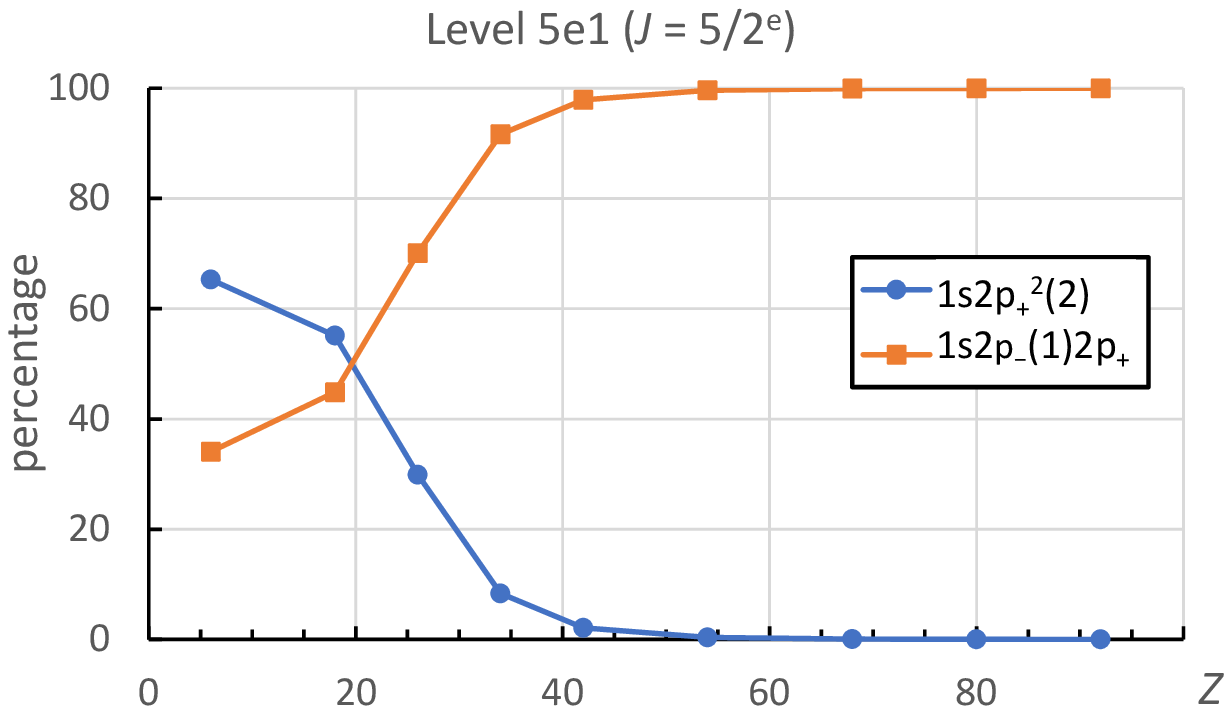}
 \includegraphics[width=.49\linewidth]{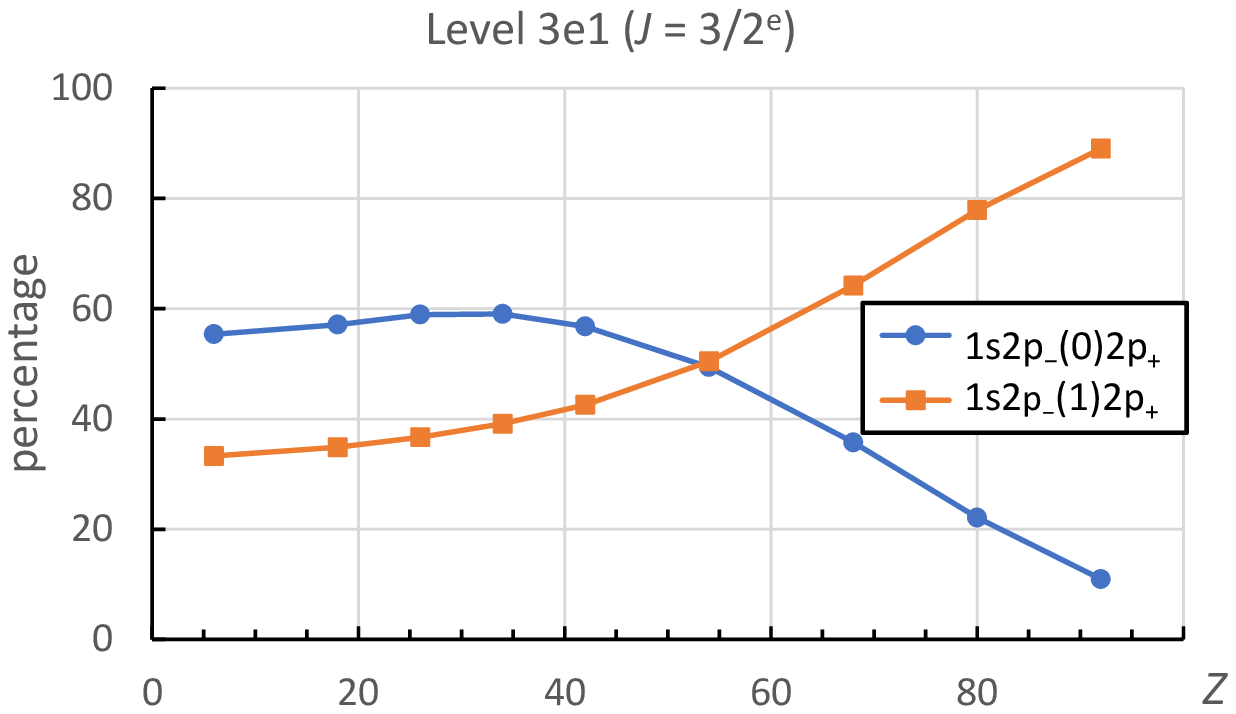}
 \includegraphics[width=.49\linewidth]{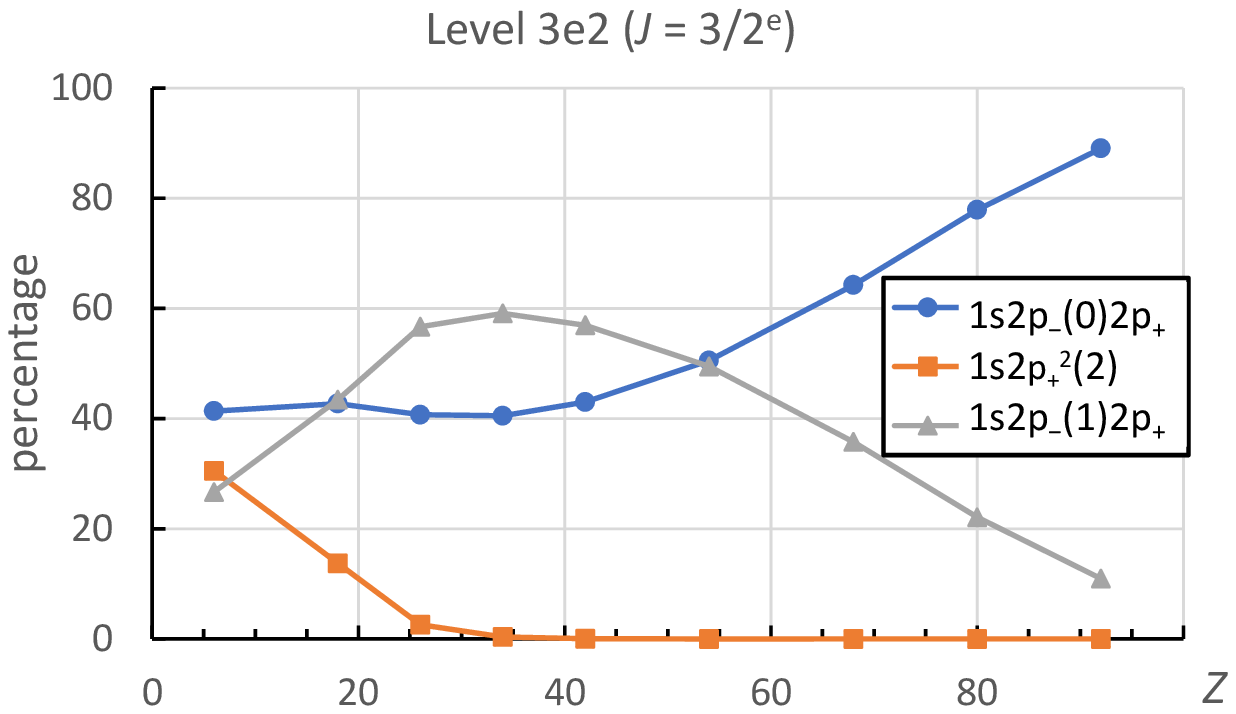}
\caption{Percentages of the eigenvector components in the composition of one 1s2s2p $J=3/2$ level (top left), one 1s2p$^2$ $J=5/2$ level (top right), and two 1s2p$^2$ $J=3/2$ levels (bottom) of Li-like ions as a function of nuclear charge $Z$. The `sequential' $jj$ coupling was used to construct the basis sets (see text).}
 \label{fig:pcnt_jj_s}
\end{figure}

Behavior of the $jj$-coupling level compositions that most drastically change along the isoelectronic sequence is shown in Fig. \ref{fig:pcnt_jj_s}. The average purity of eigenvectors (i.e., mean percentage of the leading component) in `sequential' $jj$ coupling at $Z=92$ was found to be 98~\% (odd parity) and 88~\% (even parity). Even at $Z = 34$ the $jj$ purity is higher than in the $LS$ coupling: it is 90~\% (odd) and 84~\% (even), while in $LS$ it is 89~\% (odd) and 80~\% (even). However, in the context of isoelectronic comparisons, it is not the purity that determines the best coupling scheme. It is rather the constancy of the dominant eigenvector component associated with each level. From this point of view, as can be seen in Fig. \ref{fig:pcnt_jj_s}, the $jj$ coupling has a disadvantage in the medium-$Z$ region. The two even-parity $J=3/2$ levels, 3e1 and 3e2 (bottom of Fig. \ref{fig:pcnt_jj_s}) change their leading component. For the level 3e1, this change occurs at $Z \approx 53$, while the level 3e2 changes its leading component twice: at $Z \approx 18$ and at $Z \approx 53$.   

From this point of view, the best coupling scheme describing the 1s2s2p and 1s2p$^2$ levels of Li-like ions with low and moderate $Z$ is the `reordered' $LS$ coupling (with the two $n=2$ electrons combined first to each other and then to 1s). It provides the highest average purity of level composition (100~\% for odd parity and 98~\% for even parity) at $Z = 6$. The dominant components of the eigenvectors remain the same (with percentages greater than 50~\%) up to $Z = 38$. For higher $Z$, physically more adequate level labels are provided by the `sequential' $jj$-coupling scheme, which has the highest purity of the eigenvectors at $Z=92$ (98~\% for odd parity and 88~\% for even parity) compared to all other schemes we tested. 

The level labels to be used in the ranges of $Z$ specified above are listed in Table \ref{table:labels}. 

\renewcommand{\baselinestretch}{1.0}
\begin{table}[hb!]
\caption{Level labels. $LS$ labels are valid for $Z \leq 38$, while $jj$ labels are valid for higher $Z$ (see text). 
Energy values ($E$) are quoted from Yerokhin et al. \cite{Y17,Y18} for guidance to energy ordering. The column `Lbl' contains 
labels referred to in other tables of this paper (see text).\label{table:labels}}
\begin{center}
\begin{tabular}{lllll}
\hline\\
\multirow[m]{2}{*}{Lbl} &  \multicolumn{2}{c}{$Z=6$}  & \multicolumn{2}{c}{$Z=92$} \\
 & $E$, cm$^{-1}$ \cite{Y17} & $LS$ label & $E$, cm$^{-1}$ \cite{Y18} & $jj$ label \\[10pt]
\hline\noalign{\vskip4pt}
1e0 &       0 & 1s$^2$2s $^2$S                                 & 0         & 1s$^2$2s          \\
1e1 & 2351890 & 1s2s$^2$ $^2$S                                 & 773422346 & 1s2s$^2$          \\
1e2 & 2446336 & 1s2p$^2$ $^4$P                                 & 776509696 & 1s2p$_-^2$(0)      \\
1e3 & 2481997 & 1s2p$^2$ $^2$P                                 & 810836077 & 1s2p$_-$(1)2p$_+$ \\
1e4 & 2521456 & 1s2p$^2$ $^2$S                                 & 846376043 & 1s2p$_+^2$(0)      \\
3e1 & 2446412 & 1s2p$^2$ $^4$P                                 & 810528593 & 1s2p$_-$(1)2p$_+$ \\
3e2 & 2472063 & 1s2p$^2$ $^2$D                                 & 811560453 & 1s2p$_-$(0)2p$_+$ \\
3e3 & 2482143 & 1s2p$^2$ $^2$P                                 & 846069107 & 1s2p$_+^2$(2)      \\
5e1 & 2446451 & 1s2p$^2$ $^4$P                                 & 810374741 & 1s2p$_-$(1)2p$_+$ \\
5e2 & 2471954 & 1s2p$^2$ $^2$D                                 & 845332989 & 1s2p$_+^2$(2)      \\
1o0 & 64483   & 1s$^2$2p $^2$P$^{\circ}$                       & 2264603   & 1s$^2$2p$_-$      \\
1o1 & 2371985 & 1s($^2$S)2s2p($^3$P$^{\circ}$) $^4$P$^{\circ}$ & 774051580 & 1s2s(1)2p$_-$     \\
1o2 & 2419321 & 1s($^2$S)2s2p($^3$P$^{\circ}$) $^2$P$^{\circ}$ & 776112227 & 1s2s(0)2p$_-$     \\
1o3 & 2447237 & 1s($^2$S)2s2p($^1$P$^{\circ}$) $^2$P$^{\circ}$ & 809788085 & 1s2s(1)2p$_+$     \\
3o0 & 64588   & 1s$^2$2p $^2$P$^{\circ}$                       & 35968939  & 1s$^2$2p$_+$      \\
3o1 & 2371989 & 1s($^2$S)2s2p($^3$P$^{\circ}$) $^4$P$^{\circ}$ & 773775700 & 1s2s(1)2p$_-$     \\
3o2 & 2419419 & 1s($^2$S)2s2p($^3$P$^{\circ}$) $^2$P$^{\circ}$ & 809179263 & 1s2s(1)2p$_+$     \\
3o3 & 2447226 & 1s($^2$S)2s2p($^1$P$^{\circ}$) $^2$P$^{\circ}$ & 810879642 & 1s2s(0)2p$_+$     \\
5o1 & 2372086 & 1s($^2$S)2s2p($^3$P$^{\circ}$) $^4$P$^{\circ}$ & 808211489 & 1s2s(1)2p$_+$     \\[4pt]
\hline
\end{tabular}
\end{center}
\end{table}

Throughout this paper, we use letter labels to denote the $n=2 \rightarrow n=1$ satellite transitions of Li-like ions and the nearby transitions of He-like ions. Translation of these labels to the spectroscopic designations of the lower and upper levels of the transitions is given in Table \ref{table:sat_labels}. Most of these labels were defined by Gabriel \cite{Gabriel_1972}. The forbidden transition labels E and U were introduced by Beier and Kunze \cite{Beier_1978}.

\renewcommand{\baselinestretch}{1.0}
\begin{table}[hb!]
\caption{Classification of satellite transitions near resonance transitions of He-like ions in the $LS$ and $jj$ coupling schemes. 
The first column contains transition labels, most of which were defined by Gabriel \cite{Gabriel_1972}. 
The numbers 1 and 2 in the column headings refer to the lower and upper levels of the transitions.
The columns `Lbl1' and `Lbl2' refer to the column `Lbl' of Table \ref{table:labels}. For transitions in Li-like ions, 
the $LS$-coupling labels are applicable to $Z \leq 38$, while the $jj$-coupling labels are applicable for higher $Z$ (see text).
Either $LS$- or $jj$-labels unambiguously define the levels of transitions in He-like ions in the entire sequence.\label{table:sat_labels}}
\begin{center}
\begin{tabular}{lllllll}
\hline\\
Label & Lbl1 & Lbl2 & Level1 ($LS$) & Level2 ($LS$) & Level1 ($jj$) & Level2 ($jj$) \\[10pt]
\hline\noalign{\vskip4pt}
a & 3o0 & 3e3 & 1s$^2$2p $^2$P$^{\circ}_{3/2}$ & 1s2p$^2$ $^2$P$_{3/2}$                               & (1s$^2$2p$_+$)$_{3/2}$ & [1s2p$_+^2$(2)]$_{3/2}$     \\
b & 1o0 & 3e3 & 1s$^2$2p $^2$P$^{\circ}_{1/2}$ & 1s2p$^2$ $^2$P$_{3/2}$                               & (1s$^2$2p$_-$)$_{1/2}$ & [1s2p$_+^2$(2)]$_{3/2}$     \\
c & 3o0 & 1e3 & 1s$^2$2p $^2$P$^{\circ}_{3/2}$ & 1s2p$^2$ $^2$P$_{1/2}$                               & (1s$^2$2p$_+$)$_{3/2}$ & [1s2p$_-$(1)2p$_+$]$_{1/2}$ \\
d & 1o0 & 1e3 & 1s$^2$2p $^2$P$^{\circ}_{1/2}$ & 1s2p$^2$ $^2$P$_{1/2}$                               & (1s$^2$2p$_-$)$_{1/2}$ & [1s2p$_-$(1)2p$_+$]$_{1/2}$ \\
e & 3o0 & 5e1 & 1s$^2$2p $^2$P$^{\circ}_{3/2}$ & 1s2p$^2$ $^4$P$_{5/2}$                               & (1s$^2$2p$_+$)$_{3/2}$ & [1s2p$_-$(1)2p$_+$]$_{5/2}$ \\
E & 1o0 & 5e1 & 1s$^2$2p $^2$P$^{\circ}_{1/2}$ & 1s2p$^2$ $^4$P$_{5/2}$                               & (1s$^2$2p$_-$)$_{1/2}$ & [1s2p$_-$(1)2p$_+$]$_{5/2}$ \\
f & 3o0 & 3e1 & 1s$^2$2p $^2$P$^{\circ}_{3/2}$ & 1s2p$^2$ $^4$P$_{3/2}$                               & (1s$^2$2p$_+$)$_{3/2}$ & [1s2p$_-$(1)2p$_+$]$_{3/2}$ \\
g & 1o0 & 3e1 & 1s$^2$2p $^2$P$^{\circ}_{1/2}$ & 1s2p$^2$ $^4$P$_{3/2}$                               & (1s$^2$2p$_-$)$_{1/2}$ & [1s2p$_-$(1)2p$_+$]$_{3/2}$ \\
h & 3o0 & 1e2 & 1s$^2$2p $^2$P$^{\circ}_{3/2}$ & 1s2p$^2$ $^4$P$_{1/2}$                               & (1s$^2$2p$_+$)$_{3/2}$ & [1s2p$_-^2$(0)]$_{1/2}$     \\
i & 1o0 & 1e2 & 1s$^2$2p $^2$P$^{\circ}_{1/2}$ & 1s2p$^2$ $^4$P$_{1/2}$                               & (1s$^2$2p$_-$)$_{1/2}$ & [1s2p$_-^2$(0)]$_{1/2}$     \\
j & 3o0 & 5e2 & 1s$^2$2p $^2$P$^{\circ}_{3/2}$ & 1s2p$^2$ $^2$D$_{5/2}$                               & (1s$^2$2p$_+$)$_{3/2}$ & [1s2p$_+^2$(2)]$_{5/2}$     \\
k & 1o0 & 3e2 & 1s$^2$2p $^2$P$^{\circ}_{1/2}$ & 1s2p$^2$ $^2$D$_{3/2}$                               & (1s$^2$2p$_-$)$_{1/2}$ & [1s2p$_-$(0)2p$_+$]$_{3/2}$ \\
l & 3o0 & 3e2 & 1s$^2$2p $^2$P$^{\circ}_{3/2}$ & 1s2p$^2$ $^2$D$_{3/2}$                               & (1s$^2$2p$_+$)$_{3/2}$ & (1s2p$_-$(0)2p$_+$)$_{3/2}$ \\
m & 3o0 & 1e4 & 1s$^2$2p $^2$P$^{\circ}_{3/2}$ & 1s2p$^2$ $^2$S$_{1/2}$                               & (1s$^2$2p$_+$)$_{3/2}$ & [1s2p$_+^2$(0)]$_{1/2}$     \\
n & 1o0 & 1e4 & 1s$^2$2p $^2$P$^{\circ}_{1/2}$ & 1s2p$^2$ $^2$S$_{1/2}$                               & (1s$^2$2p$_-$)$_{1/2}$ & [1s2p$_+^2$(0)]$_{1/2}$     \\
o & 3o0 & 1e1 & 1s$^2$2p $^2$P$^{\circ}_{3/2}$ & 1s2s$^2$ $^2$S$_{1/2}$                               & (1s$^2$2p$_+$)$_{3/2}$ & (1s2s$^2$)$_{1/2}$          \\
p & 1o0 & 1e1 & 1s$^2$2p $^2$P$^{\circ}_{1/2}$ & 1s2s$^2$ $^2$S$_{1/2}$                               & (1s$^2$2p$_-$)$_{1/2}$ & (1s2s$^2$)$_{1/2}$          \\
q & 1e0 & 3o2 & 1s$^2$2s $^2$S$_{1/2}$         & 1s($^2$S)2s2p($^3$P$^{\circ}$) $^2$P$^{\circ}_{3/2}$ & (1s$^2$2s)$_{1/2}$     & [1s2s(1)2p$_+$]$_{3/2}$     \\
r & 1e0 & 1o2 & 1s$^2$2s $^2$S$_{1/2}$         & 1s($^2$S)2s2p($^3$P$^{\circ}$) $^2$P$^{\circ}_{1/2}$ & (1s$^2$2s)$_{1/2}$     & [1s2s(0)2p$_-$]$_{1/2}$     \\
s & 1e0 & 3o3 & 1s$^2$2s $^2$S$_{1/2}$         & 1s($^2$S)2s2p($^1$P$^{\circ}$) $^2$P$^{\circ}_{3/2}$ & (1s$^2$2s)$_{1/2}$     & [1s2s(0)2p$_+$]$_{3/2}$     \\
t & 1e0 & 1o3 & 1s$^2$2s $^2$S$_{1/2}$         & 1s($^2$S)2s2p($^1$P$^{\circ}$) $^2$P$^{\circ}_{1/2}$ & (1s$^2$2s)$_{1/2}$     & [1s2s(1)2p$_+$]$_{1/2}$     \\
u & 1e0 & 3o1 & 1s$^2$2s $^2$S$_{1/2}$         & 1s($^2$S)2s2p($^3$P$^{\circ}$) $^4$P$^{\circ}_{3/2}$ & (1s$^2$2s)$_{1/2}$     & [1s2s(1)2p$_-$]$_{3/2}$     \\
U & 1e0 & 5o1 & 1s$^2$2s $^2$S$_{1/2}$         & 1s($^2$S)2s2p($^3$P$^{\circ}$) $^4$P$^{\circ}_{5/2}$ & (1s$^2$2s)$_{1/2}$     & [1s2s(1)2p$_+$]$_{5/2}$     \\
v & 1e0 & 1o1 & 1s$^2$2s $^2$S$_{1/2}$         & 1s($^2$S)2s2p($^3$P$^{\circ}$) $^4$P$^{\circ}_{1/2}$ & (1s$^2$2s)$_{1/2}$     & [1s2s(1)2p$_-$]$_{1/2}$     \\
w &     &     & 1s$^2$ $^1$S$_0$               & 1s2p $^1$P$^{\circ}_1$                               & (1s$^2$)$_0$           & (1s2p$_+$)$_1$              \\
x &     &     & 1s$^2$ $^1$S$_0$               & 1s2p $^3$P$^{\circ}_2$                               & (1s$^2$)$_0$           & (1s2p$_+$)$_2$              \\
y &     &     & 1s$^2$ $^1$S$_0$               & 1s2p $^3$P$^{\circ}_1$                               & (1s$^2$)$_0$           & (1s2p$_-$)$_1$              \\
z &     &     & 1s$^2$ $^1$S$_0$               & 1s2s $^3$S$_1$                                       & (1s$^2$)$_0$           & (1s2s)$_1$                  \\[4pt]
\hline
\end{tabular}
\end{center}
\end{table}

$LS$- and $jj$-coupling percentage compositions of the Li-like ion levels calculated in this work for $Z = 6$, 18, 26, 34, 42, 54, 68, 80, and 92 are given in Tables \ref{table:composition_LS} and \ref{table:composition_jj}, respectively.

\LTright=0pt
\LTleft=0pt
\footnotesize
\begin{center}
\begin{landscape}
\begin{longtable}{@{\extracolsep\fill}llllllllll}
\caption{Percentage composition of the levels of Li-like ions in the `reordered' $LS$ coupling scheme from the present calculations. The top row specifies the nuclear charge $Z$. The basis states corresponding to the labels in column Lbl are given 
in the column of $Z=6$ of Table A. The leading percentage in the compositions given for each ion corresponds to the label given in the first column, unless it is followed by a label of another basis state. Contributions of one or two other basis states are specified as well, if they are 1~\% or greater. 
\label{table:composition_LS}}

Lbl & 6 & 18 & 26 & 34 & 42 & 54 & 68 & 80 & 92 \\
\hline\noalign{\vskip3pt}
\endfirsthead
\caption[]{(continued)}
Lbl & 6 & 18 & 26 & 34 & 42 & 54 & 68 & 80 & 92 \\
\hline\noalign{\vskip3pt}
\endhead
\hline\noalign{\vskip3pt}
\multicolumn{10}{r}{\itshape (continued on next page)}\\
\endfoot
\noalign{\vskip3pt}\hline
\endlastfoot

1e0 & 100 & 100 & 100 & 100 & 100 & 100 & 100 & 100 & 100 \\
1e1 & 91 + 9 1e4 & 93 + 7 1e4 & 94 + 5 1e4 & 95 + 4 1e4 & 94 + 4 1e4 & 96 + 2 1e4 & 96 + 2 1e4 & 96 + 2 1e4 & 87 + 6 1e2 + 5 1e4 \\
1e2 & 100 & 99 & 94 + 4 1e4 & 79 + 12 1e4 + 7 1e3 & 63 + 20 1e4 + 13 1e3 & 52 + 27 1e4 + 18 1e3 & 47 + 29 1e4 + 20 1e3 & 45 + 30 1e4 + 20 1e3 & 41 + 28 1e4 + 19 1e3 \\
1e3 & 100 & 99 & 93 + 4 1e4 & 83 + 13 1e2 & 75 + 23 1e2 & 70 + 30 1e2 & 68 + 32 1e2 & 68 + 32 1e2 & 67 + 33 1e2 \\
1e4 & 91 + 9 1e1 & 92 + 7 1e1 & 87 + 5 1e3 & 80 + 10 1e3 + 8 1e2 & 74 + 13 1e2 + 12 1e3 & 71 + 17 1e2 + 12 1e3 & 69 + 20 1e2 + 12 1e3 & 68 + 21 1e2 + 11 1e3 & 67 + 21 1e2 + 11 1e3 \\
3e1 & 100 & 100 & 98 & 96 + 2 3e2 & 93 + 4 3e2 & 91 + 6 3e2 & 89 + 7 3e2 & 89 + 8 3e2 & 89 + 6 3e2 + 5 3e3 \\
3e2 & 100 & 94 + 6 3e3 & 79 + 21 3e3 & 70 + 29 3e3 & 66 + 33 3e3 & 63 + 37 3e3 & 61 + 38 3e3 & 60 + 39 3e3 & 61 + 39 3e3 \\
3e3 & 100 & 94 + 6 3e2 & 79 + 20 3e2 & 69 + 27 3e2 & 64 + 30 3e2 + 6 3e1 & 60 + 32 3e2 + 9 3e1 & 58 + 32 3e2 + 10 3e1 & 57 + 33 3e2 + 10 3e1 & 56 + 33 3e2 + 10 3e1 \\
5e1 & 100 & 99 & 87 + 13 5e2 & 62 + 38 5e2 & 53 5e2 + 47 5e1 & 61 5e2 + 39 5e1 & 64 5e2 + 36 5e1 & 65 5e2 + 35 5e1 & 65 5e2 + 35 5e1 \\
5e2 & 100 & 99 & 87 + 13 5e1 & 62 + 38 5e1 & 53 5e1 + 47 5e2 & 61 5e1 + 39 5e2 & 64 5e1 + 36 5e2 & 65 5e1 + 35 5e2 & 65 5e1 + 35 5e2 \\
1o0 & 100 & 100 & 100 & 100 & 100 & 100 & 100 & 100 & 100 \\
1o1 & 100 & 100 & 99 & 97 + 2 1o3 & 95 + 3 1o3 & 92 + 4 1o3 & 91 + 6 1o3 & 90 + 6 1o3 & 90 + 7 1o3 \\
1o2 & 100 & 100 & 98 + 2 1o3 & 90 + 10 1o3 & 83 + 16 1o3 & 78 + 22 1o3 & 75 + 24 1o3 & 75 + 25 1o3 & 74 + 25 1o3 \\
1o3 & 100 & 100 & 98 + 2 1o2 & 89 + 9 1o2 & 81 + 15 1o2 & 74 + 19 1o2 + 7 1o1 & 70 + 21 1o2 + 9 1o1 & 69 + 21 1o2 + 10 1o1 & 68 + 22 1o2 + 10 1o1 \\
3o0 & 100 & 100 & 100 & 100 & 100 & 100 & 100 & 100 & 100 \\
3o1 & 100 & 100 & 97 + 2 3o3 & 89 + 7 3o3 & 78 + 15 3o3 + 7 3o2 & 67 + 23 3o3 + 9 3o2 & 61 + 28 3o3 + 10 3o2 & 59 + 30 3o3 + 11 3o2 & 58 + 32 3o3 + 11 3o2 \\
3o2 & 100 & 93 + 6 3o3 & 82 + 16 3o3 & 66 + 24 3o3 + 10 3o1 & 52 + 28 3o3 + 20 3o1 & 39 + 31 3o1 + 30 3o3 & 38 3o1 + 34 3o3 + 28 3o2 & 40 3o1 + 34 3o3 + 26 3o2 & 42 3o1 + 35 3o3 + 24 3o2 \\
3o3 & 100 & 93 + 7 3o2 & 82 + 17 3o2 & 69 + 30 3o2 & 57 + 41 3o2 & 52 3o2 + 46 3o3 & 61 3o2 + 38 3o3 & 63 3o2 + 36 3o3 & 66 3o2 + 34 3o3 \\
5o1 & 100 & 100 & 100 & 100 & 100 & 100 & 100 & 100 & 100 \\[4pt]
\end{longtable} 

\begin{longtable}{@{\extracolsep\fill}llllllllll}
\caption{Percentage composition of the levels of Li-like ions in the `sequential' $jj$-coupling scheme from the present calculations. The basis states corresponding to the labels in column Lbl are given in the column of $Z=92$ of Table \ref{table:labels}. \label{table:composition_jj}}
Lbl & 6 & 18 & 26 & 34 & 42 & 54 & 68 & 80 & 92 \\
\hline\noalign{\vskip3pt}
\endfirsthead
\caption[]{(continued)}
Lbl & 6 & 18 & 26 & 34 & 42 & 54 & 68 & 80 & 92 \\
\hline\noalign{\vskip3pt}
\endhead
\hline\noalign{\vskip3pt}
\multicolumn{10}{r}{\itshape (continued on next page)}\\
\endfoot
\noalign{\vskip3pt}\hline
\endlastfoot
1e0 & 100 & 100 & 100 & 100 & 100 & 100 & 100 & 100 & 100 \\
1e1 & 89 + 7 1e4 + 3 1e2 & 91 + 5 1e4 + 4 1e2 & 92 + 4 1e2 + 4 1e4 & 94 + 4 1e2 + 2 1e4 & 94 + 5 1e2 + 1 1e4 & 95 + 5 1e2 & 95 + 5 1e2 & 95 + 5 1e2 & 95 + 5 1e2 \\
1e2 & 43 + 34 1e3 + 22 1e4 & 51 + 30 1e3 + 19 1e4 & 66 + 21 1e3 + 12 1e4 & 84 + 8 1e3 + 5 1e4 & 92 + 4 1e1 + 2 1e3 & 94 + 5 1e1 & 95 + 5 1e1 & 95 + 5 1e1 & 95 + 5 1e1 \\
1e3 & 65 + 23 1e2 + 12 1e4 & 69 + 25 1e2 + 6 1e4 & 78 + 20 1e2 + 1 1e4 & 91 + 8 1e2 & 97 + 2 1e2 & 100 & 100 & 100 & 100 \\
1e4 & 59 + 30 1e2 + 10 1e1 & 70 + 21 1e2 + 9 1e1 & 83 + 10 1e2 + 6 1e1 & 93 + 3 1e2 + 3 1e1 & 97 + 1 1e1 + 1 1e2 & 99 & 100 & 100 & 100 \\
3e1 & 55 3e2 + 33 3e1 + 11 3e3 & 57 3e2 + 35 3e1 + 8 3e3 & 59 3e2 + 37 3e1 + 4 3e3 & 59 3e2 + 39 3e1 + 2 3e3 & 57 3e2 + 43 3e1 + 1 3e3 & 50 + 49 3e2 & 64 + 36 3e2 & 78 + 22 3e2 & 89 + 11 3e2 \\
3e2 & 41 + 31 3e3 + 27 3e1 & 43 3e1 + 43 3e2 + 14 3e3 & 57 3e1 + 41 3e2 + 3 3e3 & 59 3e1 + 41 3e2 & 57 3e1 + 43 3e2 & 51 + 49 3e1 & 64 + 36 3e1 & 78 + 22 3e1 & 89 + 11 3e1 \\
3e3 & 58 + 39 3e1 + 2 3e2 & 78 + 22 3e1 & 93 + 7 3e1 & 98 + 2 3e1 & 99 & 100 & 100 & 100 & 100 \\
5e1 & 65 5e2 + 34 5e1 & 55 5e2 + 45 5e1 & 70 + 30 5e2 & 92 + 8 5e2 & 98 + 2 5e2 & 100 & 100 & 100 & 100 \\
5e2 & 65 5e1 + 34 5e2 & 55 5e1 + 45 5e2 & 70 + 30 5e1 & 92 + 8 5e1 & 98 + 2 5e1 & 100 & 100 & 100 & 100 \\
1o0 & 99 & 100 & 100 & 100 & 100 & 100 & 100 & 100 & 100 \\
1o1 & 88 + 11 1o3 & 91 + 9 1o3 & 94 + 6 1o3 & 97 + 3 1o3 & 98 + 1 1o3 + 1 1o2 & 98 + 2 1o2 & 97 + 3 1o2 & 95 + 5 1o2 & 93 + 7 1o2 \\
1o2 & 57 + 38 1o3 + 5 1o1 & 67 + 30 1o3 + 3 1o1 & 85 + 15 1o3 & 95 + 5 1o3 & 98 + 2 1o3 & 98 + 2 1o1 & 97 + 3 1o1 & 95 + 5 1o1 & 93 + 7 1o1 \\
1o3 & 50 + 42 1o2 + 6 1o1 & 61 + 33 1o2 + 6 1o1 & 79 + 15 1o2 + 5 1o1 & 92 + 5 1o2 + 3 1o1 & 97 + 1 1o1 + 1 1o2 & 99 & 100 & 100 & 100 \\
3o0 & 99 & 100 & 100 & 100 & 100 & 100 & 100 & 100 & 100 \\
3o1 & 55 + 44 3o2 & 62 + 38 3o2 & 73 + 26 3o2 & 87 + 13 3o2 & 95 + 5 3o2 & 99 + 1 3o2 & 100 & 100 & 100 \\
3o2 & 57 3o3 + 24 3o2 + 19 3o1 & 45 3o3 + 35 3o2 + 20 3o1 & 53 + 29 3o3 + 18 3o1 & 74 + 16 3o3 + 10 3o1 & 86 + 9 3o3 + 4 3o1 & 94 + 5 3o3 + 1 3o1 & 97 + 3 3o3 & 98 + 2 3o3 & 98 + 2 3o3 \\
3o3 & 43 + 31 3o2 + 25 3o1 & 55 + 27 3o2 + 18 3o1 & 71 + 20 3o2 + 9 3o1 & 84 + 13 3o2 + 3 3o1 & 90 + 8 3o2 + 1 3o1 & 95 & 97 + 3 3o2 & 98 + 2 3o2 & 98 + 2 3o2 \\
5o1 & 99 & 100 & 100 & 100 & 100 & 100 & 100 & 100 & 100 \\[4pt]

\end{longtable} 
\end{landscape}
\end{center}

\clearpage
\normalsize

\section{Collection and evaluation of experimental data}
\label{S:2}
Our collection of experimental data includes absolute measurements of excitation energies for the levels of 1s2$l$2$l^{\prime}$ core-excited states \cite{ASD,Aglitskii_1974,Apicella_1983,Beier_1978,Beiersdorfer_1991,Beiersdorfer_1993,Beiersdorfer_1999,Beiersdorfer_2002,Biedermann_2003,Biemont_2000,Bitter_1979,Boiko_1978,Bombarda_1988,Bruch_1979,Bruch_1987,Bruch_1991,Bryunetkin_1995,Bryzgunov_1982,Chantler_2000,Clementson_2010,Cocke_1974a,Cocke_1974b,Decaux_2003,Deschepper_1982,Dohmann_1978,Dohmann_1979,Dyakin_1997,Elton_2000,Faenov_1995,Feldman_1974,Feldman_1980,Flemberg_1942,Fu_2010,Gabriel_1969,Gatuzz_2013,Golts_1975,Gonzalez_2006,Grineva_1973,Groeneveld_1975,Gu_2005,Hell_2016,Hofmann_1990,Jannitti_1990,Kadar_1990,Kauffman_1973,Kauffman_1975,Kilgus_1993,Kononov_1977,Lee_1991,Lee_1992,Lie_1971,Machado_2020,Mack_1987,Magunov_2002,Mann_1987,Mannervik_1997,Matthews_1973,Matthews_1976,McLaughlin_2017,Mowat_1976,Muller_2009,Nicolosi_1977,Parkinson_1975,Payne_2014,Peacock_1969,Peacock_1973,Porter_2009,Pospieszczyk_1975,Rice_1995,Rice_2014,Rodbro_1979,Roth_1968,Rudolph_2013,Safronova_1977,Sawyer_1962,Schlesser_2013,Schmidt_2004,Schneider_1977,Seely_1985,Seely_1986,Smith_1995,Stolterfoht_1977,Suleiman_1994,Tarbutt_2001,TFR_Group_1982,TFR_Group_1985,TFR_Group_1985a,Thorn_2008,Walker_1971,Walker_1974,Wargelin_2001,Widmann_1995,Yao_2009}, as well as relative measurements of the fine structure and term separations from 1s2s2p $^4$P$^{\circ}$ \cite{Buchet_1984,Engstrom_1987,Knystautas_1985,Livingston_1978,Livingston_1984,Martinson_1983,Trabert_1982,Trabert_1983}. Our aim is to collect and review all available experimental data related to energy levels of the 1s2$l$2$l^{\prime}$ core-excited configurations in Li-like ions. The criteria for data to be included in our collection for comparison with theory are (i) availability of estimations of uncertainties of experimental energies and (ii) reasonably small deviations of energies obtained in the given experiment from those obtained in other experiments, as well as from theoretical predictions. Unfortunately, many studies, especially those published 40 years ago or earlier, either do not include estimations of measurement uncertainties or provide too rough estimations. For such works, we tried to estimate uncertainties by considering the number of decimals in the values published, comparing the results presented with other data obtained by the same group and/or on a similar experimental setup, by verifying the calibration standards and methods used, and/or by other methods.

Another problem we ran into is that many studies had used outdated reference data to calibrate the experimental setup. Those data, which include experimental and theoretical energies of levels and transitions, often turned out to be corrected in later publications. For such works, we adjusted the measured energies and/or corrected the related uncertainties. 

For some studies, a review of the experiment and the data obtained shows that calibration of the experimental setup was done inaccurately or not quite correctly: the number of the standards used was too small, they did not cover the entire range of the experimental results, and/or the complexity of the calibration curve did not correspond to the number and the range of standards used. Where possible, we corrected the calibration curve used in the experiment, adjusted the experimental energies and/or revised the related uncertainties of energy values. All uncertainties given in the current study are on the level of one standard deviation.

In many studies, spectral resolution was insufficient to resolve components of multiple blended transitions. In some of these studies, the centers of gravity of the unresolved blends were measured very precisely, but the separation between the unresolved components was much larger than the measurement precision. This was the case, e.g., in the work of Hell et al. \cite{Hell_2016} on the spectra of Si and S. For the Si q+r blend, the statistical uncertainty (on the 1$\sigma$ level, reduced from their reported value on the 90~\% confidence level) was 0.00011~{\AA}, while the theoretical separation between q and r is 0.00200~{\AA} \cite{Y18}. If we use the reported central wavelength to derive the upper energy levels, 1s($^2$S)2s2p($^3$P$^{\circ}$) $^2$P$^{\circ}_{1/2,3/2}$ (levels 1o2 and 3o2, see Table \ref{table:labels}) of these two transitions, they would have the same value that strongly differs from theory, giving a wrong indication of a failure of theory. One way to deal with this is to compare not the energy levels derived from experiments, but the directly measured line positions. However, this would require a derivation of the corresponding theoretical positions for centers of blended lines. This, in turn, would require knowledge of relative intensities of all blended transitions. The latter strongly depend on plasma conditions that are not precisely known in most experiments. Furthermore, it would greatly increase the number of quantities used in comparisons, as in different experiments there are different blended transitions, and decrease the statistics of data available for comparison for each quantity. All experimental data involving blended transitions are for $Z \leq 29$, where the energy structure is described by the $LS$ coupling sufficiently well (see the previous Section). Therefore, in the present work we derive the energy levels of precisely measured blended transitions by assuming the theoretical separation between fine-structure components of $LS$ terms as calculated by Yerokhin et al. \cite{Y17,Y18}. If the observed blends involve transitions from different terms, we increase the measurement uncertainties according to theoretical separations between the terms.

In Refs. \cite{Beiersdorfer_1991,Beiersdorfer_1993,Bitter_1979,Bryunetkin_1995,Bryzgunov_1982,Faenov_1995,Feldman_1980,Parkinson_1975,Seely_1985,Seely_1986,TFR_Group_1985,TFR_Group_1985a}, all experimental measurements were made relative to the resonance line w (1s$^2$~$^1$S$_0$ -- 1s2p~$^1$P$^{\circ}_1$) or intercombination line y (1s$^2$~$^1$S$_0$ -- 1s2p~$^3$P$^{\circ}_1$) of the corresponding helium-like ions. Many of these measurements also used the lines x (1s$^2$~$^1$S$_0$ -- 1s2p~$^3$P$^{\circ}_2$) and z (1s$^2$~$^1$S$_0$ -- 1s2s~$^3$S$_1$) for calibration. Reference wavelengths of these lines were taken from other experimental or theoretical studies. In 2005, after these articles had been published, Artemyev et al. \cite{Artemyev_2005} presented refined calculations of energy levels for the $n = 1$ and $n = 2$ states of He-like ions with nuclear charge in the range from $Z = 12$ to $Z = 100$. A significant improvement in precision of theoretical predictions was achieved, especially in the high-$Z$ region. For $Z < 12$, theoretical wavelengths of Yerokhin and Surzhykov \cite{Yerokhin_2019} are presently considered to be the most precise ones. They agree well with those of Drake \cite{Drake_1988} but, unlike the latter, they are accompanied by well determined small uncertainties. The calculated wavelengths of Refs. \cite{Artemyev_2005,Yerokhin_2019,Drake_1988} are in good agreement with experimental data, with the calculated uncertainties provided being generally much smaller than the experimental ones. This good agreement with a large body of experimental data was recently convincingly demonstrated by Indelicato \cite{Indelicato19}. Therefore, we adjusted experimental data from the articles quoted above to the wavelengths of the lines w, x, y, and z taken from Ref. \cite{Yerokhin_2019} for $Z < 12$ and from Ref. \cite{Artemyev_2005} for $Z \geq 12$.

In 1942, Flemberg \cite{Flemberg_1942} calibrated his curved-crystal spectrograph with characteristic X-ray lines of various chemical elements. At that time, reference wavelengths of these lines were expressed in the so-called X units (X.U.). To convert the measured wavelengths to~{\AA} units, he used the relation 1000.00~X.U. = 1.00200~{\AA}. In 1973, 31 years later, Deslattes and Henins \cite{Deslattes_1973} corrected this relation to 1000.00~X.U. = 1.0020802~{\AA}. As a result, all Flemberg’s wavelengths given in angstroms are in error by roughly a factor of 1.0020802/1.00200 = 1.00008004. We analyzed Flemberg's data for each of the elements studied (F, Mg, and Al) and determined the exact scaling factors for his X units by taking a weighted average for the particular reference lines he used, taking as standards the modern data on characteristic lines from Deslattes et al. \cite{Deslattes_2003}. These factors turned out to be 1.0020758(11) for F, 1.002073(7) for Mg, and 1.002078(6) for Al. The systematic uncertainties stemming from uncertainties of those scaling factors were taken into account in the derivation of energy levels from Flemberg's corrected data.

Aglitskii et al. \cite{Aglitskii_1974} and Boiko et al. \cite{Boiko_1978} used Flemberg’s original (uncorrected) wavelengths expressed in angstroms as standards. Therefore, their wavelengths should be scaled by roughly a factor of 1.00008004. Besides that, in both these studies the experimental wavelengths of the He-like lines w and y differ notably from the much more precise theoretical values given by Artemyev et al. \cite{Artemyev_2005} and by Yerokhin and Surzhykov \cite{Yerokhin_2019}. To take this into account, we recalibrated all their reported wavelengths using the reference data from the above sources, as well as the reference wavelengths for the H-like Ly$\alpha$ lines from Yerokhin and Shabaev \cite{Y15}. For higher members of the H-like resonance series, we used the reference data from the Atomic Spectra Database (ASD) \cite{ASD} of the National Institute of Standards and Technology (NIST).

Parkinson \cite{Parkinson_1975} reported high-resolution X-ray measurements of emission from solar active regions made with three rocket-borne crystal spectrometers. This study includes many lines of several spectra, including Ne-like Fe~XVII and Ni~XIX, as well as F-like Fe XVIII, so we found it important to analyse these data in detail. We re-calibrated Parkinson's reported wavelengths for each crystal separately by using precise theoretical wavelengths of resonance lines of H-like and He-like Ne, Na, and Mg \cite{Y15,Yerokhin_2019,ASD}. For the ADP and gypsum crystals, it was found appropriate to make additive corrections of $-0.0013(4)$~{\AA} and $+0.0012(8)$~{\AA}, respectively, where the values in parentheses (0.4~m{\AA} and 0.8~m{\AA}) represent the uncertainties of the weighted least-squares fit. For the KAP crystal, we found a linear correction amounting to $-0.0080(10)$~{\AA} at 9.177~{\AA}, $-0.0005(7)$~{\AA} at 12.51~{\AA}, $+0.0095(15)$~{\AA} at 17.041~{\AA}, and $+0.0207(27)$~{\AA} at 22.07~{\AA}. The uncertainties of the least-squares fit of this correction can be approximated by a cubic polynomial. We found that the statistical uncertainties for unblended lines are about 0.0012~{\AA} for the ADP crystal and 0.0025~{\AA} for both the gypsum and KAP crystals below 14~{\AA}. For the region between 14~{\AA} and 15.2~{\AA} recorded with the KAP crystal, the measurements are hopelessly spoiled by an inexplicable distortion of the dispersion curve. For longer wavelengths, the statistical uncertainties of the corrected KAP measurements increase from 0.003~{\AA} at (15.259--17.086)~{\AA} to 0.007~{\AA} at (18--19)~{\AA} and 0.010~{\AA} above that. For blended lines, such as the j and k lines in Li-like Mg~X, which are on a shoulder of the much stronger z line, the statistical uncertainties specified above must be doubled. For this compilation, we derived the energy of the 1s($^2$S)2s2p($^3$P$^{\circ}$) $^2$P$^{\circ}$ term of Ne~VIII from a weighted mean of the unresolved q and r satellites in the gypsum and KAP crystal spectra. For Mg~X, only the ADP crystal measurements were used, since the gypsum and KAP spectra appear to have a much lower resolution.    

One of the purposes of the work by Kononov et al. \cite{Kononov_1977} was to experimentally observe, measure, and interpret the satellite structure near the Fe~XXV resonance lines. The wavelength-measurement uncertainty of spectral lines was stated to be 0.0003~{\AA}. The lines of interest for the present study are in the range of (1.85--1.88)~{\AA}. A comparison of the measured wavelengths of the w and y lines with the values given by Artemyev et al. \cite{Artemyev_2005} showed a difference of about $-0.0006$~{\AA} and 0.0003~{\AA}, respectively. In the initial analysis, we assumed that all identifications given by Kononov et al. are correct. Under this assumption, disagreement between the values of the 1s$^2$2p~$^2$P$^{\circ}$ $J = 3/2$--1/2 fine-structure splitting derived from several pairs of observed lines prompted us to double the wavelength uncertainties. However, a closer examination showed that this disagreement is due to several incorrect identifications. Kononov et al. \cite{Kononov_1977} based their identifications on the old calculations of Vainshtein and Safronova (see references in \cite{Kononov_1977}). Those theoretical wavelengths are systematically lower than the much more precise ones calculated in Refs. \cite{Artemyev_2005,Y18} by 0.00047(9)~{\AA}. This systematic error is comparable in size with separations between the observed lines. Thus, it was easy to misidentify the lines. For example, the line observed at 1.8552~{\AA} was interpreted as the Li-like satellite s. Based on the new calculations quoted above, we have identified it as the He-like 1s$^2$ $^1$S$_0$ -- 1s2p $^3$P$^{\circ}_2$ forbidden transition (x in Gabriel's designations \cite{Gabriel_1972}), while the neighboring line at 1.8563~{\AA} has been re-interpreted as a blend of the s and m satellite transitions. Similarly, the line at 1.8678~{\AA} has been re-interpreted as a blend of the Li-like satellite l and the He-like forbidden transition z (1s$^2$ $^1$S$_0$ -- 1s2p $^3$S$_1$). Assignments of the satellite transitions q, e, and u have been moved from the lines at 1.8604~{\AA}, 1.8721~{\AA}, and 1.8730~{\AA} to those at 1.8614~{\AA}, 1.8730~{\AA}, and 1.8739~{\AA}, respectively. We also added an assignment of the satellite transition b to the previous classification of the line at 1.8580~{\AA}. After these revisions, all observed wavelength intervals appear to be statistically consistent with the wavelength uncertainty of 0.0003~{\AA} stated by Kononov et al. \cite{Kononov_1977}. However, we increased it to (0.0004--0.0006)~{\AA} for multiply classified lines where the relative intensities of the contributing transitions are not known. We have derived the Fe~XXIV energy levels from the original observed wavelengths of Kononov et al. with our revised classifications by using the least-squares level optimization code LOPT \cite{Kramida_2011} and used them in the present analysis. 

Bombarda et al. \cite{Bombarda_1988} observed the spectrum of Li-like Ni in the vicinity of the He-like resonance line w with a carefully characterized crystal spectrometer. These observations were made in emission of the JET tokamak. The relative positions of the reported lines were stated to be accurate to about 0.0001~{\AA}. By comparing the reported wavelengths of the He-like lines w, x, y, and z with the reference values of Artemyev et al. \cite{Artemyev_2005}, we determined that the reported wavelengths must be shifted by a constant correction of 0.00017(6)~{\AA}. This correction was used to derive the experimental energy levels reported here. In addition to He- and Li-like lines, Bombarda et al. reported observed wavelengths of three lines of Be-like Ni. Their corrected wavelengths are 1.60444(11)~{\AA}, 1.60884(11)~{\AA}, and 1.61084(11)~{\AA}.

In the study by Safronova and Sidelnikov \cite{Safronova_1977}, Ni spectra were investigated in the region of (1.58--1.61)~{\AA}. Uncertainties of the wavelength measurements were not specified. However, the measurements were made with the same quartz crystal spectrometer and with the same calibration method as in Ref. \cite{Kononov_1977}, and thus the same uncertainty of 0.0003~{\AA} can be expected for unblended lines. We analyzed the data of Ref.~\cite{Safronova_1977} by comparing them with theoretical values of Artemyev et al. \cite{Artemyev_2005} and of Yerokhin and Surzhykov \cite{Y18} in the same way as described above for Ref. \cite{Kononov_1977}. In addition, we used the wavelengths of three lines of Be-like Ni observed by Bombarda et al. \cite{Bombarda_1988} and corrected by us as described above. It turned out that many of the line classifications given in Ref. \cite{Safronova_1977} must be revised. In particular, the three Be-like lines mentioned above were mis-identified, and the line assigned to the g satellite must have been blended with the He-like z line. The revisions of the He-like satellite assignments are straightforward, as they are based on matching the observed wavelengths with the theoretical ones from the above references. By comparing the reported wavelengths of the He-like w and y lines with the reference theoretical values \cite{Artemyev_2005} and the wavelengths of the three Be-like lines with the more precise experimental values described above, we determined that there is a significant systematic shift well described by a quadratic polynomial. To derive the Ni~XXVI energy levels from the data of Ref.~ \cite{Safronova_1977}, we used our revised classifications and removed the systematic shift. Its uncertainty was added in quadrature to the statistical uncertainties. 

In the study of He- and Li-like Mo by Beier and Kunze \cite{Beier_1978}, an ingenious technique of noise suppression by analyzing correlations of intensities recorded in several spectrograms was used. The uncertainties of the reported wavelengths were specified as $\pm$0.0002~{\AA}. Beier and Kunze stated that theoretical wavelengths they used to identify the observed lines are accurate to $\pm$0.0005~{\AA}. They referred to Ivanov et al. \cite{Ivanov_1975}. However, that paper was about He-like ions only. For He-like Mo, indeed, the reference wavelengths used by Beier and Kunze agree with those of Ref. \cite{Artemyev_2005} within $\pm$0.0003~{\AA}. However, for Li-like Mo, the theoretical wavelengths of Goldsmith \cite{Goldsmith_1974} (which were used by Beier and Kunze) differ from those of Yerokhin and Surzhykov \cite{Y18} by $\pm$0.003~{\AA} on average. As a result, some of the line identifications of Beier and Kunze had to be revised. We dropped both classifications of the line observed at 0.6859~{\AA}, as their new theoretical wavelengths are too far from the observed one. Assignment of the r satellite transition has been moved from the line observed at 0.6885~{\AA} to 0.6928~{\AA}. Assignment of the s satellite transition has been dropped from the line at 0.6893~{\AA}. Assignments of the t, q, and U satellite transitions have been added to the multiple classifications of the lines at 0.6885~{\AA}, 0.6893~{\AA}, and 0.6912~{\AA}, respectively. The precise positions of the lines depicted in Figure 4 of Beier and Kunze \cite{Beier_1978} were determined by digitizing this figure and establishing a quadratic dispersion correction by using the He-like w, x, y, and z as reference standards. The resulting wavelengths were found to be accurate to better than 0.00010~{\AA} (statistical uncertainty), while the systematic uncertainty due to the limited precision of calibration varied between 0.00003~{\AA} and 0.00010~{\AA}. These experimental data are the only ones currently available for Li-like Mo.

Unlike the experiment described above, we did not find it possible to utilize any of the results of the subsequent study of Beier \cite{Beier_1979} on Li-like Nb. The peaks observed were too weak to allow a confident detection of lines.

The X-ray measurements of the He-like and Li-like Ar spectra performed with the beam-foil method by Dohmann and Mann in 1979 \cite{Dohmann_1979} using a crystal spectrometer were made with a very high resolution, extraordinary for this type of experiments. The full width at half intensity of the narrowest lines was only 1.4~m{\AA}, which can be seen in Figure 2 of their paper. This shows a great potential of the now unpopular beam-foil technique. The observed wavelengths listed in their Table 1 were stated to have relative uncertainties of 0.3~m{\AA}, while the absolute uncertainty was 0.8~m{\AA}. The large systematic error was due to the lack of precise wavelength standards available at that time. Imprecision of the earlier calculations of wavelengths of the Li-like satellite lines led to misidentification of several observed lines. Nevertheless, the data of Dohmann and Mann were found to be usable after re-calibration with internal standards provided by high-precision calculations of Yerokhin and Surzhykov \cite{Yerokhin_2019} for the He-like w line, measurements of Machado et al. \cite{Machado_2020} for the Li-like q, r, and U lines, and measurements of Tarbutt et al. \cite{Tarbutt_2001} for the Li-like m line. Fitting the differences between the wavelengths reported by Dohmann and Mann and the reference wavelengths from the above sources revealed a systematic error varying linearly with wavelength. It amounted to $-0.57$~m{\AA} at the w line and $+0.73$~m{\AA} at the U line. The magnitude of this error is consistent with the absolute uncertainty stated by Dohmann and Mann. The satellite lines n, o, and p were found to be misidentified and were excluded from the present compilation. The line designated as q was found to be an unresolved blend of the q, b, and r transitions, while the line designated as r was re-classified as a blend of the a and d transitions. The line originally assigned to the a+d blend was discarded as spurious. The line assigned to the j satellite was also discarded. Similarly, the line assigned to the v satellite was re-classified as the u transition, while the line designated as h was re-classified as a blend of the v and h transitions. After these revisions and correction of the calibration, all observed wavelengths are in good agreement with other observations, as well as with the calculations of Yerokhin and Surzhykov \cite{Y18}. It should be noted that the spectrogram shown in Figure 2 of Dohmann and Mann \cite{Dohmann_1979} contains many unidentified lines tentatively interpreted as transitions in Be-like, B-like, and C-like Ar. Their identification remains a challenge for theorists and experimentalists.

Bi{\'e}mont et al. \cite{Biemont_2000} reported observed wavelengths in the inner-shell transition spectra of Li-like through F-like argon recorded at the PF-1000 plasma focus facility in Warsaw, Poland. They stated that their spectrograms were calibrated with an uncertainty of 0.3~m{\AA}. However, the reported wavelengths of the n and m Li-like satellite lines differ from the much more precise measurements of Tarbutt et al. \cite{Tarbutt_2001} by about 5~m{\AA}. It seems probable that there was a large systematic error in the wavelength scale calibration of Bi{\'e}mont et al. caused by an error in determining the observed position of the He-like resonance line w. That line was one of the six reference lines used for calibration. As seen in Fig. 6 of Bi{\'e}mont et al. \cite{Biemont_2000}, this line (denoted as R in the figure) is broad and asymmetric, which was probably caused by blending with $n \geq 3$ Li-like satellite transitions. We re-calibrated the portion of the spectrum related to the observed Li-like transitions by using the nine wavelengths measured by Tarbutt et al. \cite{Tarbutt_2001} as references. A quadratic polynomial was used in this correction. The corrected wavelengths agree with those of Tarbutt et al. within 1~m{\AA} on average, which we adopted as the uncertainty. For the very weak satellite line n (peak 1 on Fig. 6 of Bi{\'e}mont et al. \cite{Biemont_2000}), the uncertainty was doubled.

In the article by Beiersdorfer et al. \cite{Beiersdorfer_1999} describing measurements of nitrogen spectra made on an electron beam ion trap (EBIT) with a grazing incidence spectrometer, the w, y and z lines of nitrogen were used to verify the calibration of the experimental setup. Observed wavelengths of these lines were compared with the values calculated by Drake \cite{Drake_1988}, and a very good agreement was noted in the article. For instance, for the w line, the experimental wavelength is 28.7861(22)~{\AA}, while the calculated value is 28.7870~{\AA}. The authors stated that the wavelength of the q line at 29.4135(37)~{\AA} obtained in the experiment agrees well with the value of 29.443~{\AA} calculated by Vainshtein and Safronova in 1978 \cite{Vainshtein_1978} and with the value of 29.41~{\AA} given by Gabriel and Jordan in 1969 \cite{Gabriel_1969}. The authors did not compare their results with the existing experimental data obtained with better precision. For example, in 1977, Nicolosi and Tondello \cite{Nicolosi_1977} measured the q line at 29.434(3)~{\AA}. The value given by Yerokhin et al. \cite{Y17} for this line is 29.43006(11)~{\AA}, which is very close to the experimental value of Nicolosi and Tondello. Given the fact that the result of Beiersdorfer et al. deviates from the values of Refs. \cite{Nicolosi_1977,Y17} by a factor of 4.3 and 4.5 times the combined uncertainties, respectively, we tried to find a cause of the difference. Having analyzed Figure 3 in the article, we noticed some discrepancy. Our measurements of line positions on the published plot gave wavelengths of the lines w, y and z within 0.006~{\AA} from the published values. Our determination of the q line position gave a wavelength of 29.431(6)~{\AA}, which differs significantly (by 0.0175~{\AA}) from the published value at 29.4135~{\AA}. The q line position corresponding to the published value is visibly off the plotted peak. As privately communicated to us by P. Beiersdorfer, the spectrum tracing depicted in Figure 3 of Ref. \cite{Beiersdorfer_1999} corresponds to one of the seven independent recordings used to derive the published wavelength value. The latter was obtained as an average of all these seven measurements, and there were no misprints in the published average value. The large deviation from the mean for the one measurement presented in the Figure discussed above indicates that there was a large scatter in the seven measurements of the q line, which was not taken into account in derivation of the small uncertainty given in Ref. \cite{Beiersdorfer_1999}. Since the original measurement data are not available, it is impossible for us to correct the published value. Thus, we have omitted it from the present compilation.

In the EBIT measurement of He- and Li-like X-ray spectra of vanadium \cite{Beiersdorfer_1991}, made with a Si crystal spectrometer, wavelength calibration was made with the H-like Ly$\alpha$ and He-like w lines. While the reference wavelengths they used for the Ly$\alpha_{1,2}$ lines are sufficiently close to the presently recommended values of Yerokhin and Shabaev \cite{Y15}, the wavelength they chose to use for the w line was lower than that of Refs. \cite{Drake_1988,Artemyev_2005,Yerokhin_2019} by 0.00008~{\AA}. This choice was grounded on previously observed systematic discrepancies of a few experimental measurements in isoelectronic spectra with theory. As noted in the beginning of this Section, the very large body of experimental data accumulated since then refuted this discrepancy \cite{Indelicato19}. Thus, we re-calibrated the reported Li-like wavelengths by using the Ly$\alpha_{1,2}$, w, x, y, and z lines as standards, with the wavelengths taken from Refs. \cite{Y15,Artemyev_2005}. For the y line, we assumed blending with the hyperfine-induced 1s$^2$~$^1$S$_0$--1s2p~$^3$P$^{\circ}_0$ transition with the same intensity ratio of 4:1 as used by Beiersdorfer et al. \cite{Beiersdorfer_1991}. We assumed the observed wavelengths of the Ly$\alpha_{1,2}$ and w to be equal to those used by Beiersdorfer et al. as standards with uncertainties of 0.00005~{\AA}, twice smaller than the smallest uncertainty reported for other lines. The correction to the dispersion curve was assumed to be a linear function of wavelength. 

The work of Chanter et al. \cite{Chantler_2000} was devoted to measurements of lines in He-like vanadium. We inferred their observed wavelength of the q satellite in Li-like vanadium from their Figure 5 using the precise wavelengths of the w, x, y, and z lines \cite{Artemyev_2005} as standards.

In the work of Nicolosi and Tondello \cite{Nicolosi_1977} mentioned above, the wavelengths of H-like and Li-like dielectronic satellites near H-like and He-like resonance lines of Be~II, B~III, C~IV, N~V, and O~VI were measured in laser-produced plasmas using a 600~lines/mm 2-m grazing incidence grating spectrometer. They used the wavelengths of H- and He-like resonance lines as calibration standards. To verify their results, we compared their measured wavelengths of the He-like w and `x,y' lines (in C~IV, N~V, and O~VI) with the modern precisely calculated values from Yerokhin and Surzhykov \cite{Yerokhin_2019}. The wavelengths of the w lines were given in the captions of Figures 7, 8, and 9 of Ref. \cite{Nicolosi_1977} with limited precision. We restored the more precise values 40.270(7)~{\AA}, 28.795(10)~{\AA}, and 21.602(6)~{\AA} for C~V, N~VI, and O~VII, respectively, from their Figure 11. This revealed a problem with their C~V and C~IV measurements. Namely, their value for the wavelength of the `x,y' blend strongly deviates from the theoretical wavelength of the intercombination line y \cite{Yerokhin_2019}. The forbidden line x (1s$^2$~$^1$S$_0$--1s2p~$^3$P$_2$) is expected to give a negligibly small contribution to the intensity of the `x,y' blend in laser-produced plasmas and thus to determination of its wavelength. To investigate the origin of the disagreement, we digitized Figure 7 of Nicolosi and Tondello \cite{Nicolosi_1977} and fitted their measured spectrum with Gaussian peak profiles. It turned out that the `x,y' line in carbon is blended on the short-wavelength side with an equally strong pair of satellites m and n, which resulted in the wavelength disagreement mentioned above. We also found that the wavelength 40.816(10)~{\AA} given by Nicolosi and Tondello for the `s,t' line of C~IV corresponds to a misidentified peak of unknown origin, while the somewhat weaker `s,t' line is partially resolved from it on the long-wavelength side. By decomposing this blended profile into two Gaussian peaks, we determined the wavelength of the `s,t' line to be 40.857(10)~{\AA} in good agreement with the theoretical value, 40.86250(10)~{\AA} \cite{Yerokhin_2019}, where the uncertainty is entirely due to the unknown relative intensities of the s and t transitions. Furthermore, the wavelength 41.341(10)~{\AA} given by Nicolosi and Tondello for the `q,r' line of C~IV turned out to correspond to the center of gravity of an unresolved blend of the q, r, a, b, c, and d transitions. Therefore, we did not include the measurements of the `s,t' and `q,r' lines of C~IV from Ref. \cite{Nicolosi_1977} in this compilation. On the other hand, all their results for N~V and O~VI were confirmed.

Roth and Elton \cite{Roth_1968} reported measurements of 208 spectral lines of C~V--VI, N~VI--VII, O~IV-VIII, and Si~VII--XII in a theta pinch made with a grazing incidence grating spectrometer in the wavelength range of (16--210)~{\AA}. Of these lines, about half were measured only in the first diffraction order, while the other half were also measured in the second order, and a few tens were also measured in the third and fourth orders. By comparing their observed wavelengths of H-like and He-like resonance and intercombination lines of C and O with precise reference values from \cite{Y15,Yerokhin_2019,ASD}, we determined that the measurement uncertainty in the first diffraction order was about 0.004~{\AA} for the shortest wavelengths and increased to about 0.009~{\AA} for the longest wavelengths. These values pertain to isolated lines, i.e., those that are not perturbed by blending with lines of other species, other transitions in the same species, or lines from other diffraction orders. For lines measured in several orders of diffraction, we determined the wavelength as a weighted average of all measurements. Because of the large number of species in the plasma and presence of several orders of diffraction, uncertainties of several lines had to be increased.

In the article by K{\'a}d{\'a}r et al. \cite{Kadar_1990}, Ne~VIII spectra were studied. Having the same uncertainty of 0.1~eV as in the work by Bruch et al. \cite{Bruch_1991}, the level energies of K{\'a}d{\'a}r et al. are all shifted by about 0.2~eV compared to the values given by Bruch et al., which are consistent with those given by Wargelin et al. \cite{Wargelin_2001}, Parkinson \cite{Parkinson_1975} and Stolterfoht et al. \cite{Stolterfoht_1977}. Taking this into account, as well as the fact that K{\'a}d{\'a}r et al. used a linear correction curve with a predetermined slope and only one reference point for determining the intercept, we concluded that the calibration curve used in their study was imprecise and made our adjustment of the curve and the resulting energies. The level energies have been shifted by 1623~cm$^{-1}$, and the shift uncertainty of 259~cm$^{-1}$ has been added in quadrature to the original level uncertainties. The published 1s($^2$S)2s2p($^1$P$^{\circ}$)~$^2$P$^{\circ}$ term value at 672.70(4)~eV (7354138(868)~cm$^{-1}$) does not fit the positions of levels  predicted by Yerokhin et al. \cite{Y17} at 7371805(49)~cm$^{-1}$ and 7372047(45)~cm$^{-1}$, for $J$ = 1/2 and 3/2 respectively (levels 1o3 and 3o3 in Table \ref{table:labels}), and is far from other experiments. For the 1s($^2$S)2s2p($^1$P$^{\circ}$)~$^2$P$^{\circ}$ term, Wargelin et al. \cite{Wargelin_2001}, Stolterfoht et al. \cite{Stolterfoht_1977}, Matthews et al. \cite{Matthews_1976} and Pospieszczyk \cite{Pospieszczyk_1975} gave values  7374100(1600)~cm$^{-1}$, 7364600(5600)~cm$^{-1}$, 7369000(12000)~cm$^{-1}$, 7369000(11000)~cm$^{-1}$, respectively. Therefore, we have discarded the value of K{\'a}d{\'a}r et al. \cite{Kadar_1990} for this term as incorrectly identified. We assigned the Auger electron peak observed at 674.65(0.10)~eV (corrected as described above), which was listed as unidentified in the article by K{\'a}d{\'a}r et al., to the 1s($^2$S)2s2p($^1$P$^{\circ}$)~$^2$P$^{\circ}$ term. The measurements by K{\'a}d{\'a}r et al. \cite{Kadar_1990} were used in the study by Kramida and Buchet-Poulizac \cite{Kramida_2006}, as well as in the NIST ASD \cite{ASD}. We did not include the Ne~VIII data from these two sources in the current compilation. 

In 2013, Al~Shorman et al. \cite{Al_Shorman_2013} measured the energies of the 1s$^2$2s~$^2$S -- 1s($^2$S)2s2p($^3$P$^{\circ}$)~$^2$P$^{\circ}$ and 1s$^2$2s~$^2$S -- 1s($^2$S)2s2p($^1$P$^{\circ}$)~$^2$P$^{\circ}$ absorption resonance transitions in Li-like nitrogen at 421.472(30)~eV and 425.449(30)~eV, respectively. Deviations of these energies from theoretical values of Yerokhin et al. \cite{Y17} are 4 to 6 times larger than the declared experimental uncertainty of 0.030~eV. In 2017, McLaughlin et al. \cite{McLaughlin_2017} measured the energies of the resonance transitions 1s$^2$2s~$^2$S -- 1s($^2$S)2s2p($^3$P$^{\circ}$)~$^2$P$^{\circ}$ and 1s$^2$2s~$^2$S -- 1s($^2$S)2s2p($^1$P$^{\circ}$)~$^2$P$^{\circ}$ in Li-like oxygen at 562.940~eV and 567.620(50)~eV, where the uncertainty of the experimental energy was given relative to the first line. The additional calibration uncertainty was specified as 0.15~eV. Both experiments, by Al Shorman et al. \cite{Al_Shorman_2013} and by McLaughlin et al. \cite{McLaughlin_2017}, were performed at the SOLEIL synchrotron radiation facility in Saint-Aubin, France. It turned out that from these two similar studies in close energy regions, the work that was performed four years later than the first one has five times worse precision of the photon beam energy calibration. This caused us to doubt the small energy uncertainty given by Al Shorman et al. \cite{Al_Shorman_2013} and request those authors to re-evaluate their measurements. As privately communicated to us by them \cite{Cubanyes_2021}, the revised values for the energies of the 1s$^2$2s~$^2$S -- 1s($^2$S)2s2p($^3$P$^{\circ}$)~$^2$P$^{\circ}$ and 1s$^2$2s~$^2$S -- 1s($^2$S)2s2p($^1$P$^{\circ}$)~$^2$P$^{\circ}$ transitions in N~V are 421.39(7)~eV and 425.37(9)~eV, respectively. The corrected energy splitting of these two lines, 3.98(3)~eV, reported in the same communication, was ignored in the present work. It disagrees with the calculated difference between the centers of gravity of the unresolved upper terms of these two transitions, 4.056(6) \cite{Y17}, where the uncertainty in parentheses corresponds to a quarter of the fine-structure interval of the 1s($^2$S)2s2p($^3$P$^{\circ}$)~$^2$P$^{\circ}$ term. Although ratios of oscillator strengths of unresolved transitions from the fine-structure levels of both $^2$P$^{\circ}$ terms should be almost equal to ratios of statistical weights \cite{Chen_1986}, we assumed a large uncertainty in these predicted properties. Even so, the discrepancy of 0.08(3)~eV between the observed and predicted line separations calls for a repeated measurement.

Measurements of the wavelength of the q+r line in Li-like oxygen were reported in 15 studies \cite{Matthews_1973,Bruch_1979,Bruch_1987,Mack_1987,Hofmann_1990,Roth_1968,Gabriel_1969,Pospieszczyk_1975,Nicolosi_1977,Lee_1992,Schmidt_2004,Yao_2009,Gatuzz_2013,Liao_2013,Gatuzz_2013}. The most precise measurement was the one made on an EBIT by Schmidt et al. \cite{Schmidt_2004}. Three measurements were made using X-ray absorption spectra of interstellar media recorded by the Chandra orbital observatory \cite{Yao_2009,Gatuzz_2013,Liao_2013}. While the first two of them are in fair agreement with all other observations, the measurement of Liao et al. \cite{Liao_2013} deviates from the mean by 3.5$\sigma$, where $\sigma=0.0023$~{\AA} is the wavelength measurement uncertainty reported by Liao et al. Nevertheless, we retained this measurement in the tables for the purpose of testing our statistical analysis procedures. This will be discussed in the following Section. 

Rudolph et al. \cite{Rudolph_2013} measured the photoabsorption X-ray spectrum of He-like through F-like iron with a very high resolution using a double-crystal high-heat-load monochromator (HHLM) equipped with two Si crystal pairs. They calibrated the absolute energy scale of the HHLM with the K-edge energies of Mn, Fe, Co, Ni, and Cu directly measured by Kraft et al. \cite{Kraft_1996} with uncertainties of $\pm 20$~meV. The systematic uncertainty of this calibration was specified as $\approx 70$~meV and was stated to be strictly additive, i.e., that the relative positions of the measured lines were not affected by this systematic uncertainty. Those relative positions were claimed to be precise to between 3~meV and 6~meV for the He- and Li-like lines. However, the reported difference between the energies of the He-like w and y lines differs from the more precise value of Artemyev et al. \cite{Artemyev_2005} by 21(6)~meV, where the uncertainty in parentheses is a combination in quadrature of experimental statistical uncertainties of Rudolph et al. and total uncertainties of Artemyev et al. Furthermore, we have recently been informed that there is a problem with the design of angular encoders commonly used in measurements on synchrotrons. These devices are used to scan the photon energy in steps. As reported by Crespo L{\'o}pez-Urrutia and Leutenegger (private communication, 2021), the sizes of these steps have been found to possess periodic errors that greatly exceed manufacturers' specifications typically trusted by experimentalists. These errors, although not detected nor evaluated by Rudolph et al., were likely contained within the $\pm$70~meV systematic errors. However, they made the size of these systematic errors not constant but quasi-random. Therefore, in the present work, the statistical uncertainties assigned to the Rudolph et al. measurements were determined as combinations in quadrature of the statistical and systematic parts specified by them. The measured energies of the w and y lines reported by Rudolph et al. are greater than the reference values of Artemyev et al. \cite{Artemyev_2005} by $114(70)$~meV and $92(69)$~meV, respectively. The energies of the Be-like lines reported at 6628.804(68)~eV and 6597.858(67)~eV deviate from those previously measured by Beiersdorfer et al. \cite{Beiersdorfer_1993} also in the same direction, by $86(240)$~meV and $813(1060)$~eV. Therefore, we decreased all energies reported by Rudolph et al. by 104(48)~meV, as determined by the weighted average of these four deviations. The uncertainty of this correction, 48~meV, was treated as an additional systematic error and added in quadrature to produce the adopted total uncertainties of the corrected values. We note that the discrepancy between the predicted line separations and those observed in the synchrotron measurements of Al Shorman et al. \cite{Al_Shorman_2013} discussed above may be caused by the same quasi-random systematic effects found in the work of Rudolph et al. 

Tarbutt et al. \cite{Tarbutt_2001} measured wavelengths of the satellite lines near the $n = 2$ resonance lines of helium-like argon. In the experiment, the s and t transitions were partially blended with each other and with the y line at 3.96891(20)~{\AA}. As described by Tarbutt et al., the u and v transitions were blended with each other at 4.01012(20)~{\AA}. A linear dispersion of the spectrometer was assumed, and the wavelength calibration was done using only two lines: x and z. The w line was not used, because it did not appear on the detector simultaneously with the dielectronic satellites being studied, and the y line was not used because it was blended with the s and t lines. We did not use the wavelength of the s+t blend reported by Tarbutt et al. because the blend is on a shoulder of the several times stronger y line and because its observed value depends on the relative intensities of the two transitions that are difficult to estimate. The line observed at 4.01012(20)~{\AA} \cite{Tarbutt_2001} and ascribed to the blend of the u and v transitions is probably due to a transition in Be-like Ar (see below).

In the article by Biedermann et al. \cite{Biedermann_2003}, helium-like argon resonance lines and satellite transitions were investigated. For the observed dielectronic satellites, wavelengths were determined by calibrating the spectrum to the theoretical values of the w and z lines calculated by Drake \cite{Drake_1988} (3.94907~{\AA} and 3.99415~{\AA}, respectively). As stated in the article, one of the factors contributing to the uncertainty is non-linearity of the interpolation between the w and z lines. In the experiment, the $n = 2$ dielectronic satellite transitions of lithium-like argon were observed in the region of (3.96--4.02)~{\AA} and measured with an uncertainty of (0.0002--0.0005)~{\AA}. The wavelengths of the s, t, u, v lines reported by Biedermann et al. deviate from those of Yerokhin and Surzhykov \cite{Y18} by a factor of 3, 5, 8, and 11 times the experimental uncertainty, respectively. We made a closer analysis of the data of Biedermann et al. by comparison with the values given by Yerokhin and Surzhykov \cite{Y18} and by adjusting the calibration curve. This resulted in the increased estimated wavelength uncertainties in the range of (0.0004--0.0008)~{\AA}, which were used in our compilation. We did not include lines s and t in our compilation as having questionable wavelength values or/and uncertainties. We noticed that on the scatter plot in Figure 1 of \cite{Biedermann_2003} there is a weaker line next to the line at 4.01084(40)~{\AA} identified as a blend of u and v lines. This weaker line was not measured in the study. Our measurements of the scatter plot gave the position of this line at 4.01515(40)~{\AA}, which is very close to the positions of the u and v lines predicted by Yerokhin and Surzhykov \cite{Y18} at 4.015003(16)~{\AA} and 4.016111(16)~{\AA}, respectively. We included these new measurements of the u and v lines in the compilation. We note that the line observed at about 4.010~{\AA} was incorrectly identified as a blend of the u and v lines in Refs. \cite{Tarbutt_2001, Biedermann_2003}. This line is probably due to the 1s$^2$2s$^2$~$^1$S$_0$ -- 1s2s$^2$2p~$^1$P$_1$ transition in Be-like Ar measured by Schlesser et al. \cite{Schlesser_2013} at 4.0101287(39)~{\AA} and by Machado et al. \cite{Machado_2018} at 4.0101273(113)~{\AA}.

In the Auger electron studies of Mack and Niehaus \cite{Mack_1987} and of Lee et al. \cite{Lee_1991,Lee_1992}, energies of autoionizing doubly excited 1s2$l$2$l^{\prime}$ states 
of Li-like ions were reported for C~IV, N~V, O~VI, and F~VII in the region of (227--605)~eV. In these studies, the relative energy scale in each ion's spectrum was calibrated relative to the theoretical center of gravity of the corresponding 1s2s2p~$^4$P$^{\circ}$ peak. Those theoretical values were taken from Refs. \cite{Chen_1986,Chung_1984,Vainshtein_1978,Holoien_1967,Bruch_1987,Can_1982}. Mack and Niehaus also measured the separations between the C~IV, N~V, and O~VI 1s2s2p~$^4$P$^{\circ}$ peaks and thus were able to establish a common relative energy scale for all three ions, which was then fixed to the theoretical $^4$P$^{\circ}$ value for C~IV taken from Bruch et al. \cite{Bruch_1985}. In Table \ref{table:2} of Mack and Niehaus \cite{Mack_1987}, uncertainties are provided for all peaks, except for the $^4$P$^{\circ}$ peaks that were used in the calibration, which means that these uncertainties were given for the separations from $^4$P$^{\circ}$. From Figures 1 and 2 of their article, it is seen that the lines used for calibration are blended (not fully resolved) and asymmetric. Thus, there were systematic errors in the fitted separations resulting from the errors in the fitted position of the reference peak. Similar errors are present in the measurements of Lee et al. \cite{Lee_1991,Lee_1992}. To bring those measurements to an absolute energy scale, we adopted a two-step procedure. First, a correction was calculated for the separations from $^4$P$^{\circ}$ using the weighted average values available from all other experimental data on these spectra. Then the resulting energy levels were re-calibrated by calculating weighted average differences from the experimental means of all other studies. The systematic uncertainties of each of these two corrections were added in quadrature to the statistical uncertainties of each peak measurement given in the tables of Refs.  \cite{Mack_1987,Lee_1991,Lee_1992}. In our tables, the values referred to these studies are the ones corrected as described above.

Gonz{\'a}lez Mart{\'i}nez et al. \cite{Gonzalez_2006} observed six peaks corresponding to dielectronic recombination resonances in Li-like Hg$^{77+}$ recombining to He-like Hg$^{78+}$. They designated these peaks as He$_1$ through He$_6$. We derived excitation energies of the corresponding energy levels of Li-like Hg$^{77+}$ from their observed resonance energies by adding the ionization energy of Hg$^{77+}$ precisely calculated by Sapirstein and Cheng \cite{Sapirstein_2011}. To verify the classification of the observed energy levels, we calculated the energies along with radiative and autoionization rates using the FAC code of Gu \cite{Gu_2008}. The classifications of Gonz{\'a}lez Mart{\'i}nez et al. have been confirmed, except for the He$_5$ peak. In Ref. \cite{Gonzalez_2006}, this peak was assigned to the same energy level as the He$_3$ peak. Although the observed energies of these two peaks are very close, 48844~eV and 48845~eV, they were observed in different areas of the two-dimensional map displayed in Figure 2 of Ref. \cite{Gonzalez_2006}, meaning that the observed energies of photons emitted during radiative decay of these two resonances were significantly different, with the difference corresponding to the excitation energy of the 1s$^2$2p$_+$ level, 2370~eV \cite{Y18}. Note that in the discussion of the spectrum of Li-like Hg$^{77+}$ we use the $jj$-coupling labels for the levels, as explained in Section \ref{section:coupling}. The [1s2s(0)2p$_+$]$_{3/2}$ level (a.k.a. 3o3, see Table \ref{table:labels}) correctly assigned to the He$_3$ resonance has no allowed radiative transition to 1s$^2$2p$_+$, so the He$_5$ peak cannot originate from the same autoionizing level. However, there is another autoionizing level nearby, [1s2p$_-$(1)2p$_+$]$_{5/2}$ (a.k.a. 5e1), which has an allowed transition to 1s$^2$2p$_+$. According to our calculations, the strength of the dielectronic capture resonance from this even-parity level should be similar to that of the He$_3$ peak. In fact, very similar data were calculated by the authors of Ref. \cite{Gonzalez_2006}. Results of those calculations are listed in Table A.1 of the thesis of Gonz{\'a}lez Mart{\'i}nez \cite{Gonzalez_2005}. Thus, in the present tables, we assigned the He$_5$ peak to the [1s2p$_-$(1)2p$_+$]$_{5/2}$ level. We note that the energy level observed by Gonz{\'a}lez Mart{\'i}nez et al. \cite{Gonzalez_2006} at 48844~eV was misidentified by Yerokhin and Surzhykov \cite{Y18}. They denoted it as 1s2p$^2$ $^4$P$_{3/2}$, which corresponds to a $jj$-coupling label [1s2p$_-$(1)2p$_+$]$_{3/2}$. The predicted intensity of radiative decay from this $J=3/2$ level is much smaller than from the $J=5/2$ level, while the energies of these two levels are very close. The absolute calibration of the energy scale was made in Ref. \cite{Gonzalez_2006} with an uncertainty of ${\pm}14$~eV, while uncertainties of the energy intervals relative to the 1s2s$^2$ resonance were stated to be much smaller, between 4~eV and 9~eV. In our tables, we assigned to these data the total uncertainties equal to a combination in quadrature of the statistical and systematic contributions. All Li-like data of Gonz{\'a}lez Mart{\'i}nez et al. agree with calculations of Yerokhin and Surzhykov \cite{Y18} well within these total uncertainties. 

The only reported observations of spectra of Li-like Ga, Ge, and Y \cite{Aglitskii_1984,Aglitsky_1988,Seely_2018} do not contain numerical data, but present tracings of the recorded spectrograms. We extracted the data from the figures of Refs. \cite{Aglitskii_1984,Aglitsky_1988} by using the theoretical wavelengths of He-like lines \cite{Artemyev_1999} and reference data on X-ray characteristic lines \cite{Deslattes_2003}, while for the Ga spectrum of Seely et al. \cite{Seely_2018} we had the original raw spectrograms in a digital format. We have not included these data in the present tables because of their low precision, but they all agree with predictions of Yerokhin and Surzhykov \cite{Y18} within uncertainties. 

To conclude this Section, we note that reports of preliminary measurements that were refined and published subsequently are excluded from the present compilation. An example is the work of Indelicato et al. \cite{Indelicato_2007}, which was a preliminary result of the work published later by Schlesser et al. \cite{Schlesser_2013}.

\subsection{Beam-foil measurements of fine-structure intervals in quartet terms}
\label{S:2.1}
The beam-foil measurements of Livingston and Berry \cite{Livingston_1978}, Tr\"abert et al. \cite{Trabert_1982,Trabert_1983}, Martinson et al. \cite{Martinson_1983}, Buchet et al. \cite{Buchet_1984}, Livingston et al. \cite{Livingston_1984}, Knystautas and Druetta \cite{Knystautas_1985}, and Engstr\"om et al. \cite{Engstrom_1987} stand apart from the absolute measurements discussed above. In these experiments, only transitions between fine-structure levels of the even and odd quartet terms of ions with $Z \leq 13$ were observed, but transitions connecting these levels to the ground level were in a much shorter wavelength range not covered by the instruments used. Thus, only relative positions of the levels within the quartet system were established, making it necessary to use a different approach in comparison of these data with the calculations of Yerokhin et al. \cite{Y17}. 

In each of these beam-foil studies, wavelengths of seven fine-structure transitions (some of them blended) were measured. These data are available for nuclear charges $Z = 6$, 7, 8, 9, 10, 12, and 13. For each $Z$, wherever more than one measurement was available, we determined the weighted mean experimental wavelength, which has a slightly reduced uncertainty compared to the original measurements. From these mean observed wavelengths, one can derive the energy levels by fixing one of the quartet levels at the theoretical value of Yerokhin et al. \cite{Y17}. This preserves the experimentally determined separations between the quartet levels, which can be directly compared with the calculations. By means of a least-squares level optimization with the LOPT code \cite{Kramida_2011}, we derived optimized energy levels, Ritz wavenumbers for the fine-structure intervals, and separations between the centers of gravity (cg) of the quartet terms. Comparison of these data with theory will be discussed in Section \ref{S:3} below. These experimental data are used to derive the recommended values of energy levels and transition wavenumbers for the quartet levels of these Li-like ions, which will also be discussed in Section \ref{S:3}.

\section{Statistical analysis procedures}
\label{S:stat}
A weighted mean $M$ of several measurements $M_i (i = 1, ..., N)$ is determined using reciprocal squares of the standard measurement uncertainties $u_i$ as weights. The standard uncertainty of the weighted mean is determined as $u_M = (\Sigma u_i^{-2})^{-1/2}$. It is not uncommon to have one or more measurements significantly deviating from the weighted mean. Here, a deviation is significant when it exceeds the measurement uncertainty. Presence of such outlying measurements requires a modification of this standard procedure, because it usually indicates presence of errors that were not anticipated in those outlying measurements. This means that the stated standard uncertainty of those measurements was underestimated, which influences the value of the weighted mean and leads to underestimation of its uncertainty. Many different solutions to this problem were used in scientific literature. 

For example, in the least-squares level optimization code LOPT \cite{Kramida_2011}, the measure of uncertainty of the relative positions of energy levels is calculated using a formula formula developed by Radziemski and Kaufman \cite{Radziemski_1969}, given here in an adapted form: 

\begin{equation}
\label{eq:Radz}
u_M = \frac{[\sum{(u_i^{-2} + u_i^{-4}{\Delta}_i^2)}]^{1/2}}{\sum{u_i^{-2}}} ,
\end{equation} 
where ${\Delta}_i$ is the difference of the $i$-th measurement from the weighted mean.

This ad-hoc equation does increase the uncertainty of the weighted mean if outlying measurements are present. However, it does not change the mean value, which is equivalent to multiplying all participating uncertainties by the same factor that is greater than one. Equation (\ref{eq:Radz}) is not justified by rigorous statistical theory. In the studies where it was applied, including Ref. \cite{Kramida_2011}, its use was motivated by the argument that those unaccounted errors empirically accounted for by this equation are of systematic nature, and there cannot be any rigorous theory for systematic errors. However, the currently prevailing view of statisticians is that systematic errors of quasi-random nature should be treated in the same way as statistical errors. There are many statistical theories relevant to this question; see, e.g., the review by Rukhin \cite{Rukhin_2009} and references therein. The term `dark uncertainty' is now commonly used by statisticians for the unaccounted sources of measurement errors mentioned above. One of the best estimators of the dark uncertainty equally distributed between all participating measurements can be calculated with the Mandel--Paule algorithm (see Ref. \cite{Rukhin_2009}). Recently published works of Rukhin \cite{Rukhin_2019,Rukhin_2019a} represent an important new development in statistical theory. Namely, the method suggested in these studies accounts for the heterogeneous character of the measurements. The studies quoted above were developed for statistical treatment of interlaboratory studies. Such studies are usually performed by different people in different experimental settings, using different methods. It is counterintuitive to expect that the unanticipated measurement errors (dark uncertainty) would be distributed equally between such measurements. It is more natural to expect that such errors are present in only a limited subset of the measurements. 

Rukhin's methods \cite{Rukhin_2019,Rukhin_2019a} use the maximum likelihood principle to determine the proper division of the entire set of measurements into two clusters, `homogeneous' and `heterogeneous'. The dark uncertainty is assigned only to the members of the heterogeneous cluster. The measurements considered in the present paper are not always made in different laboratories or by different people. Nevertheless, they are all made with different experimental equipment or different light sources. For example, Lawrence Livermore National Laboratory (LLNL) has several EBIT devices and many different spectrometers, including dispersive crystals of various materials and various geometry and grating instruments. We treat measurements made with different instrumentation as independent, even though they may have been made by the same researchers.

We have implemented the two methods developed by Rukhin \cite{Rukhin_2019,Rukhin_2019a}, originally coded in the R language, in Excel Visual Basic functions within the newly developed statistical toolbox for atomic spectroscopy. The methods are called `clustered maximum likelihood estimator' (CMLE) and `clustered restricted maximum likelihood estimator' (CRMLE). In all cases where there were three or more measurements of the same energy level, the two estimators gave similar results, i.e., the same cluster division and close values of the dark uncertainty. In these cases, we used the CMLE results to determine the total uncertainties of the participating measurements and their weighted mean. However, in many cases where only two measurements are available, the CMLE and CRMLE methods resulted in alternate assignments of the `heterogeneous cluster', i.e., they pinpointed a different culprit in the case of disagreeing measurements. In such cases, we used the Mandel--Paule method (also included in our Excel toolbox) that assigns the same dark uncertainty to both measurements.

To verify the statistical consistency of various measurement results, we use the normal probability plots described in the NIST/SEMATECH Handbook of Statistical Methods \cite{Filliben_2013}. Their use will be illustrated in the following Sections.

\section{Comparison of experimental data with theoretical calculations}
\label{S:3}
\subsection{Comparison of level intervals within the quartet term system}
\label{S:3.1}
\begin{figure}[ht!]
 \centering
 \includegraphics[width=.49\linewidth]{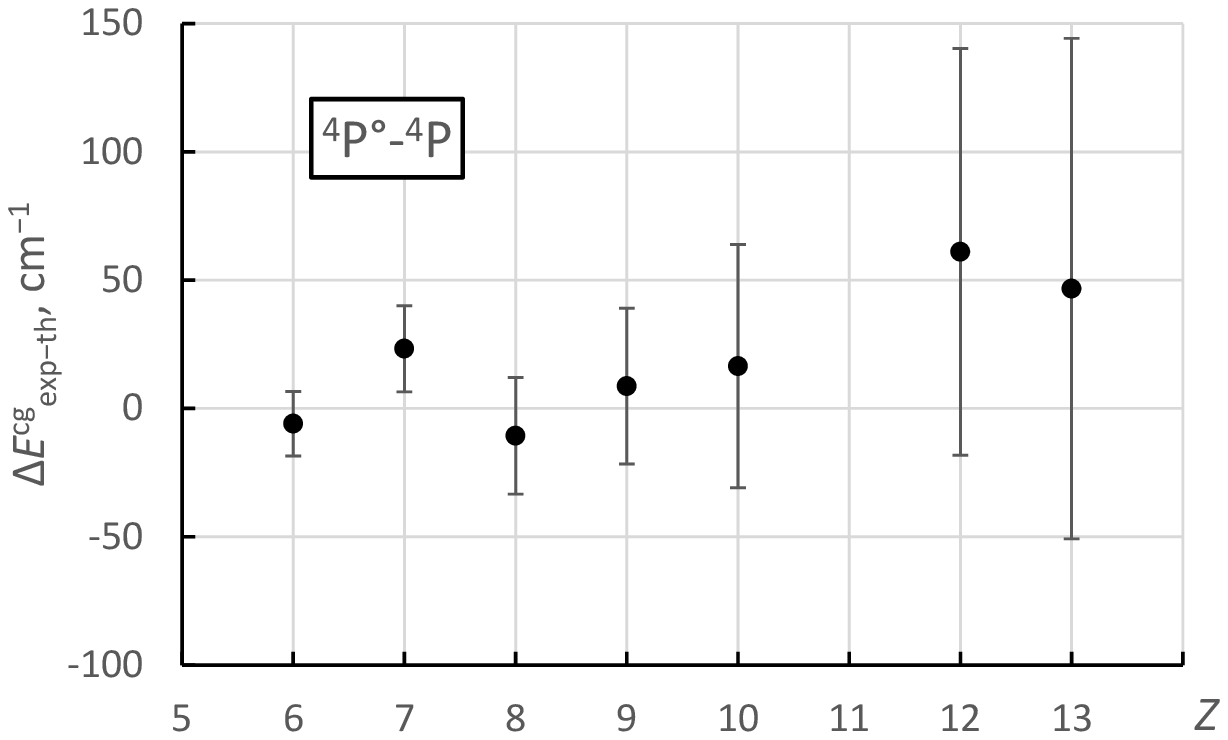}
 \includegraphics[width=.49\linewidth]{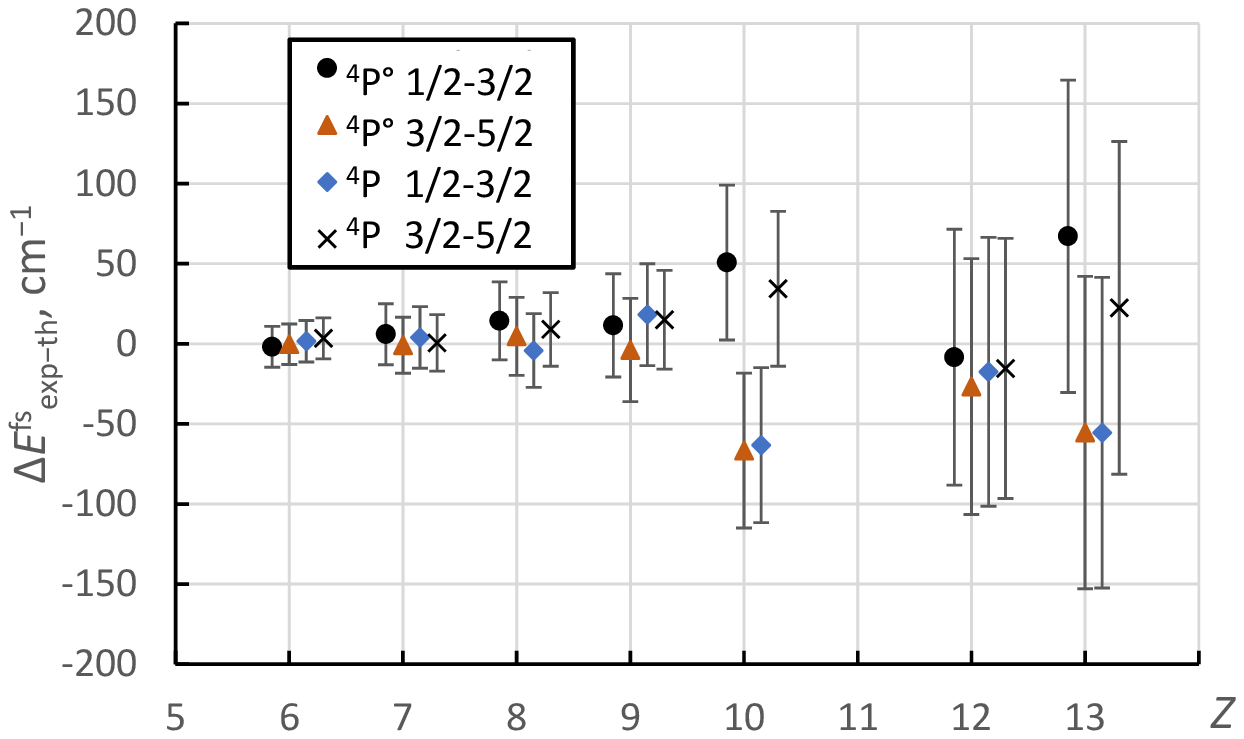} 
 \caption{Comparison of experimental and theoretical energies of the 1s($^2$S)2s2p($^3$P$^{\circ}$) $^4$P$^{\circ}$ and 1s2p$^2$ $^4$P fine-structure energy levels of Li-like ions. Left: separations between centers of gravity of the quartet terms; right: fine-structure intervals within the quartet terms. Experimental data are from beam-foil experiments (see text). Theoretical data are from Yerokhin et al. \cite{Y17}. The error bars are combinations in quadrature of the experimental and theoretical uncertainties. They are dominated by theoretical uncertainties.}
 \label{fig:bf_cmp_e_th}
\end{figure}

We started by comparing the measurements of fine-structure level separations from the levels of the 1s2s2p $^4$P$^{\circ}$ term \cite{Buchet_1984,Engstrom_1987,Knystautas_1985,Lee_1992,Livingston_1978,Livingston_1984,Mack_1987,Martinson_1983,Trabert_1982,Trabert_1983} with theoretical calculations by Yerokhin et al. \cite{Y17}. For the purpose of comparing experiment with theory, we used the method outlined in Section \ref{S:2} for deriving the energy levels from the observed wavelengths. Namely, in each ion's spectrum we fixed one quartet energy level, 1s($^2$S)2s2p($^3$P$^{\circ}$) $^4$P$^{\circ}_{3/2}$ (1o1 in Table \ref{table:labels}), at its theoretical value of Yerokhin et al. This preserves the directly observed level separations and allows them to be compared with theory. The differences between the observed and theoretical quartet energy levels are shown in Figure \ref{fig:bf_cmp_e_th}.

As seen from Figure \ref{fig:bf_cmp_e_th}, the theoretical data of Yerokhin et al. \cite{Y17} are entirely corroborated by this comparison: both the separations between the quartet terms and the fine-structure intervals within the terms agree with experiments. The uncertainties of all quantities compares in Figure \ref{fig:bf_cmp_e_th} are dominated by the theoretical uncertainties specified by Yerokhin et al., and this comparison essentially validates these uncertainty estimates.  
For the separations between the quartet terms, the beam-foil experimental data are more precise than the theory \cite{Y17} by a factor ranging between 2.5 and 7 (4.1 on average). For the fine-stricture intervals, these ratios are between 1.4 and 5.5 (2.9 on average). On the other hand, the connection between the quartet and doublet systems cannot be tested by these experimental data. Such tests are provided by the comparisons of absolute energy-level measurements described in the next subsection. They also confirm the validity of the data of Yerokhin et al. \cite{Y17}. However, experimental precision of these absolute measurements is much lower than that of the theory (by a factor ranging from 18 to 250). Therefore, we chose the theoretical values of Yerokhin et al. \cite{Y17} for the 1s($^2$S)2s2p($^3$P$^{\circ}$) $^4$P$^{\circ}_{3/2}$ base level (1o1) in each spectrum, which is used to determine the excitation energies from the ground level.

The experimental and theoretical data for the quartet levels discussed above are presented in Table \ref{table:quart_E}, where for each level we give the nuclear charge number of the atom, $Z$, the calculated energy and its uncertainty, $E_{\text{th}}$ and $u_{\text{th}}$, respectively, (in cm$^{-1}$) as given by Yerokhin et al. \cite{Y17}, recommended experimental energy $E_{\text{exp}}$ and its uncertainty $u_{\text{rel}}$ (in cm$^{-1}$) for separation from the 1s($^2$S)2s2p($^3$P$^{\circ}$) $^4$P$^{\circ}_{3/2}$ base level (1o1), and the list of references to the sources used to derive the level value. To determine the total uncertainty of the recommended energy for excitation from the ground level, the given $u_{\text{rel}}$ value must be combined in quadrature with the uncertainty $u_{\text{th}}$ of the theoretical energy of the base level. The contribution of the latter dominates in those total uncertainties.

\subsection{Comparison of absolute energy measurements with theory} \label{S:3.2} 

Absolute measurements of excitation energies are compared with theoretical calculations of Yerokhin et al. \cite{Y17,Y18} in Table \ref{table:2}. This table contains 999 measurements for $Z$ from 6 to 80. The table is divided into 16 sections with each section containing measurements of the same energy level in various spectra. The sections are sorted in the order of increasing excitation energy in C~IV. The level identifiers are given at the top of each section. They start with the (2$J$)$^{\pi}$\# identifier defined in Table \ref{table:labels} and are followed by the $LS$- and $jj$-coupling labels. As discussed in Section \ref{section:coupling}, the $LS$-coupling labels are valid for $Z \leq 38$, while the $jj$-labels should be used for greater $Z$. As mentioned in Section \ref{section:coupling}, the $jj$-labels of levels 3e1 and 3e2 for $Z \leq 53$ represent the second leading $jj$-component of their eigenvectors (see Table \ref{table:composition_jj}).

Each section of Table \ref{table:2} contains the same set of columns. The first column is the nuclear charge of the element, $Z$. The second column is the sequential number $N_m$ of the measurement for the same level and the same $Z$. Sorting of these measurements is arbitrary. The only purpose of $N_m$ is to visualize the groups of measurements of the same quantity. The total number of distinct measured quantities (energy levels of a certain ion) is 329. Since there are 999 measurements in total, the average number of measurements per quantity is about three. The largest number of measurements, fifteen, is for the levels 1o2 and 3o2 (1s($^2$S)2s2p($^3$P$^{\circ}$) $^2$P$^{\circ}_{1/2,3/2}$) in O~VI. Fourteen distinct energy levels in $Z$ = 6, 8, 10, 18, and 26 have ten or more measurements included in the table. For 137 quantities (about 42~\% of the total 329), there is only one measurement available. 

The third and fourth columns in Table \ref{table:2} give the theoretical energy and its uncertainty ($E_{\text{th}}$ and $u_{\text{th}}$), both quoted from Yerokhin et al. \cite{Y17} for $Z \leq 17$ and from Yerokhin and Surzhykov \cite{Y18} for higher $Z$. The fifth and sixth columns give the experimentally measured energy and its uncertainty, respectively ($E_{\text{exp}}$ and $u_{\text{exp}}$). The reference to the original measurement is given in the last column of the table. We remind the reader that most of these measurements have been adjusted or corrected as described in Section \ref{S:2}, while some are our new determinations based on the data or figures of the original papers. Even in the cases where no changes were made to the originally reported data, we have evaluated and validated these original data and their uncertainties. Therefore, when quoting the data from our Table \ref{table:2}, it would be prudent to include a reference to the present paper. 

The seventh column of Table \ref{table:2} gives the value of the `dark uncertainty',  assigned to each measurement by our statistical analysis procedure described in Section \ref{S:stat}. Out of 862 measurements that are not unique for the measured quantity (an energy level in a certain ion), only sixteen got assigned a non-zero dark uncertainty. Five of these dark uncertainty values are relatively small ($\leq60$~\% of the measurement uncertainty). They correspond to the cases of only a relatively small disagreement between the two or more measurements of the same quantity. The remaining cases represent a large disagreement between different studies and will be discussed further below.

The total experimental uncertainty, $u_{\text{etot}}$, given in the eighth column of Table \ref{table:2} is the sum in quadrature of $u_{\text{exp}}$ and $u_{\text{dark}}$. It was used to determine the weighted mean experimental energy, $E_{\text{em}}$, and its uncertainty, $u_{\text{em}}$, given in the ninth and tenth columns, respectively. This derivation was made with the standard statistical formulas, $E_{\text{em}}$ = $(\Sigma E_{\text{exp}}u_{\text{etot}}^{-2})/\Sigma u_{\text{etot}}^{-2}$ and $u_{\text{em}}$ = $(\Sigma u_{\text{etot}}^{-2})^{-1/2}$. To make comparisons easier, the same values of $E_{\text{em}}$ and $u_{\text{em}}$ are repeated for all measurements of the same quantity.

The eleventh column of Table \ref{table:2} contains the ratios of the uncertainties of the mean experimental and theoretical values, 
$u_{\text{em}}/u_{\text{th}}$. These ratios are given only in the rows containing the first measurement of a quantity ($N_{\text{m}} = 1$). Only seven of these ratios are smaller than 1.0 (ranging from 0.14 to 0.77). They represent the cases where the mean experimental energy $E_{\text{em}}$ is more precise than the theoretical energy $E_{\text{em}}$. These few experimental results are all in good agreement with the theory (well within the combined uncertainties) and provide our recommended values for the level 1s($^2$S)2s2p($^3$P$^{\circ}$) $^4$P$^{\circ}_{5/2}$ (5o1) in S~XIV, as well as the levels 1s($^2$S)2s2p($^3$P$^{\circ}$) $^2$P$^{\circ}_{1/2,3/2}$ (1o2 and 3o2) in S~XIV, Cl~XV, and Ar~XVI. 
The next three columns of Table \ref{table:2} contain three different normalized residuals: $R_{\text{e}-\text{em}}$, $R_{\text{em}-\text{t}}$, and $R_{\text{e}-\text{t}}$. All three represent the ratios of energy differences divided by the uncertainty of that difference. The first normalized residual, $R_{\text{e}-\text{em}}$, is for the differences between individual measurements of a quantity ($E_{\text{exp}}$) and its weighted experimental mean ($E_{\text{em}}$). These are non-blank only for the measurements that are not unique for the quantity measured and are used in the discussion of statistics of measurements. The second normalized residual, $R_{\text{em}-\text{t}}$, is used for comparing the mean experimental energy with theory. It is non-blank only in the rows with $N_{\text{m}} = 1$. The third residual, $R_{\text{e}-\text{em}}$, is given for each measurement as a measure of deviation of this measurement from theory.

The last column of Table \ref{table:2} gives the reference to each original measurement. Some of the original studies contained two or more measurements of the same spectrum. If these measurements were made with a similar technique, we derived a mean value that is given in Table \ref{table:2}. However, the results presented by Matthews et al. \cite{Matthews_1976} were obtained with two very different techniques, X-ray emission and Auger electron spectroscopy. We treat these results as independent measurements. Thus, the reference \cite{Matthews_1976} is followed by `(Auger)' or `(X-ray)' in Table 
\ref{table:2} to distinguish between these two sets of results.


For rounding of values given with uncertainties in Table \ref{table:2}, the “rule-of-99” based on the uncertainty value was used. Literally, the rule is formulated as follows: in units of the last significant digit of the value, the uncertainty must be not greater than 99 and not smaller than 10, always keeping two significant digits in the uncertainty. This preserves a reasonable numerical precision of the values in line with recommendations of International Committee for Weights and Measures (CIPM; formerly BIPM), see Ref. \cite{GUM2008}, Section 7.2.6.

In the table, the fractional number of measurements with $R_{\text{e}-\text{t}} \leq 1.0$ and $R_{\text{e}-\text{t}} \leq 2.0$ are 90.6~\% and 99.7~\%, respectively. For the ratio $R_{\text{em}-\text{t}}$, the corresponding fractions are 84.5~\% and 99.1~\%, respectively. 

\begin{figure}[ht!]
 \centering
 \includegraphics[width=.49\linewidth]{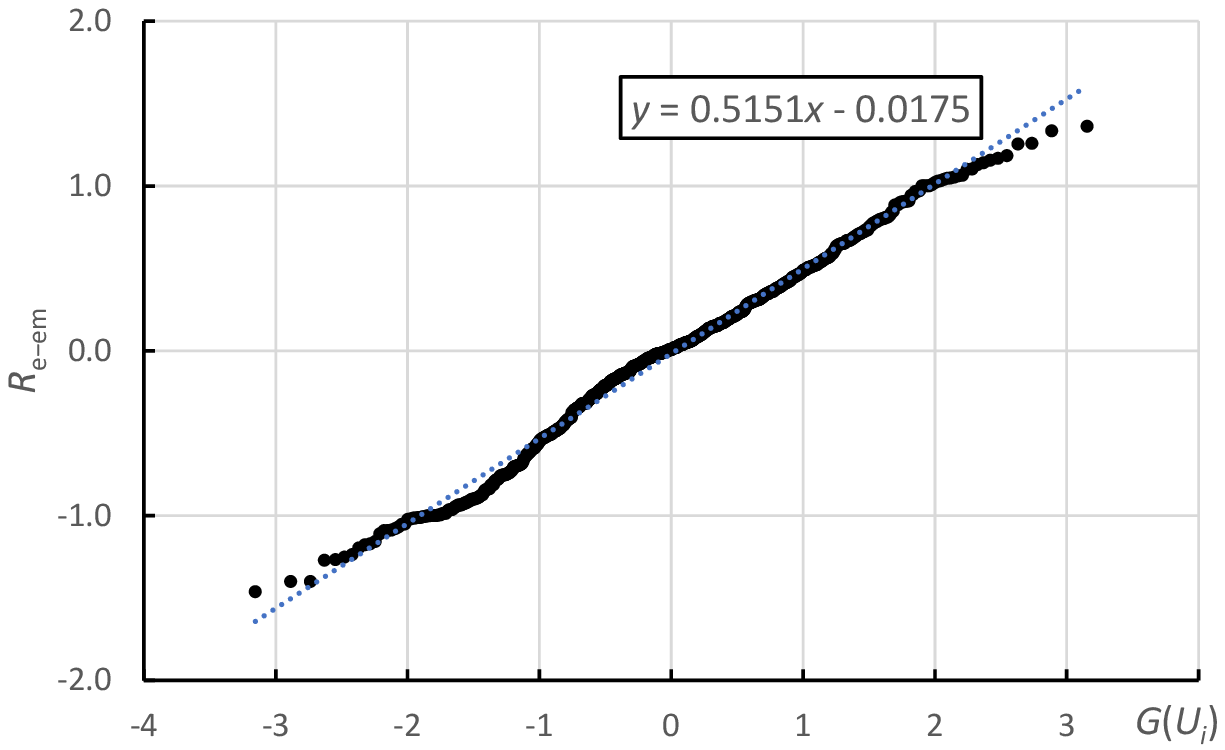}
 \includegraphics[width=.49\linewidth]{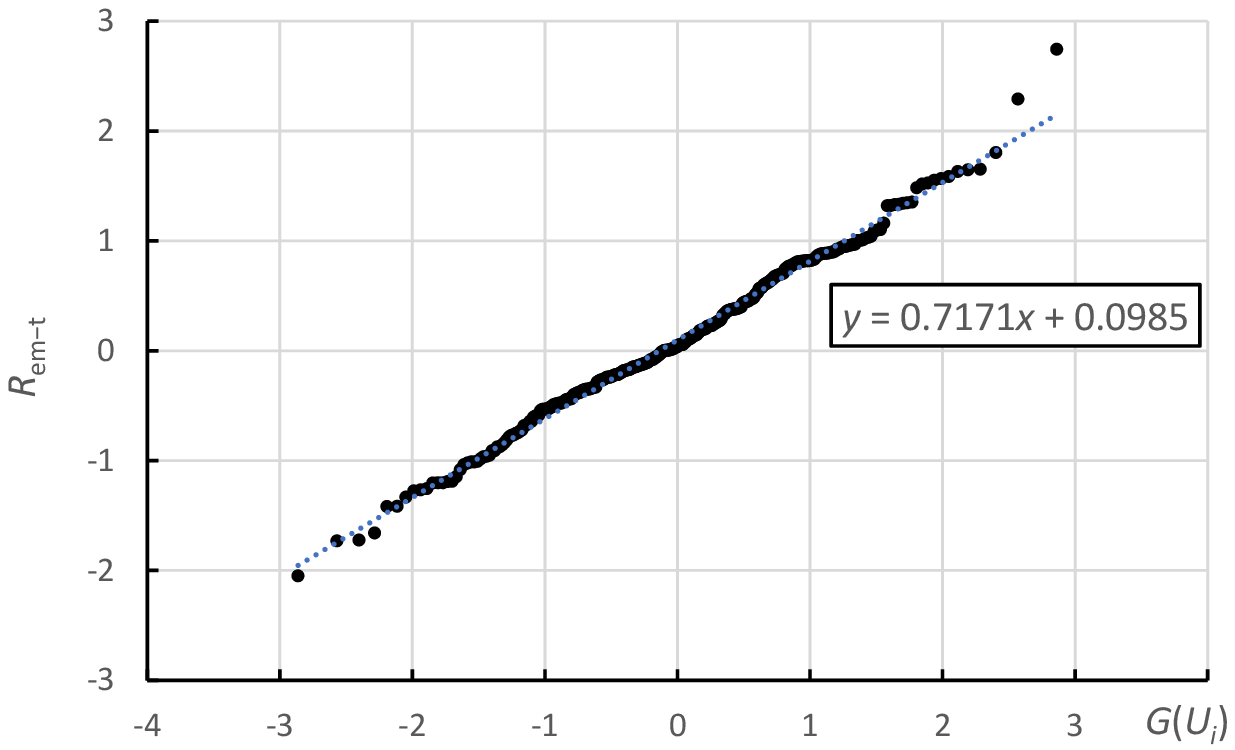}
 \includegraphics[width=.49\linewidth]{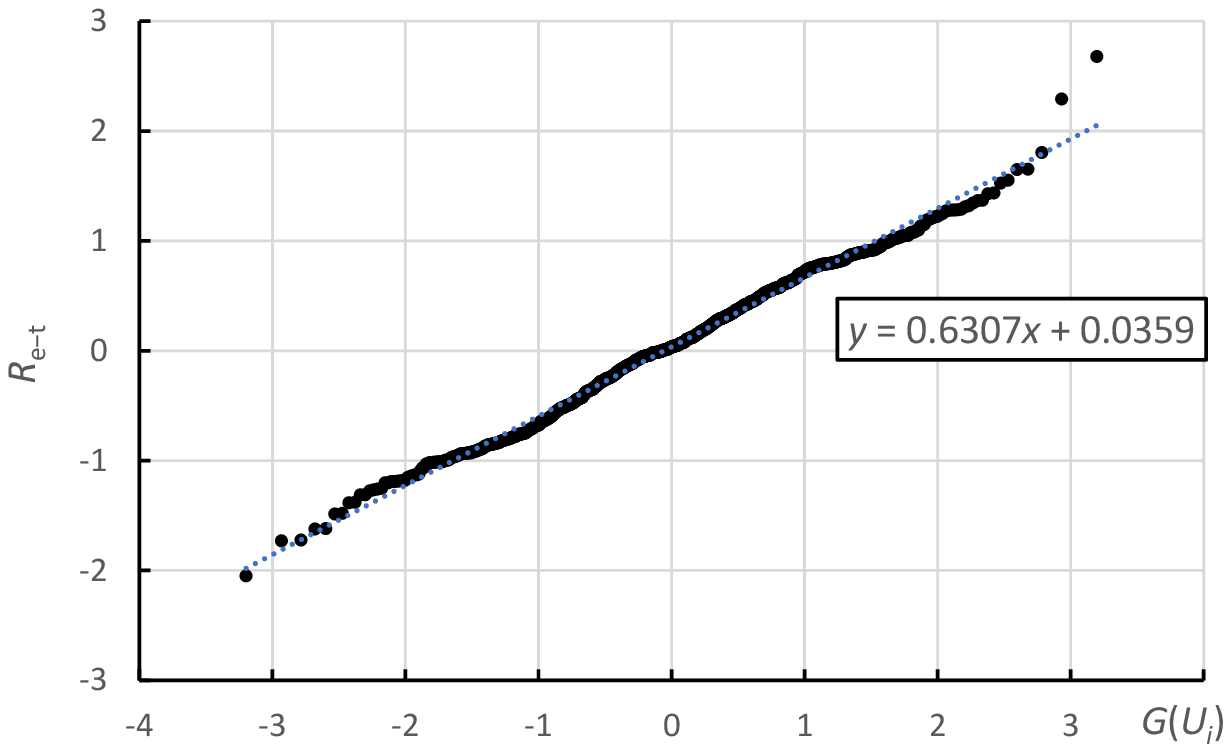}
 \caption{Normal probability plots of three normalized residuals given in Table \ref{table:2}: $R_{\text{e}-\text{em}}$ (top left), $R_{\text{em}-\text{t}}$ (top right), and $R_{\text{e}-\text{t}}$ (bottom). The quantity $G(U_i)$ on the horizontal axes of these plots is the percent point function of the uniform order statistic median of the normal distribution \cite{Filliben_2013}. The dotted lines are linear fits to the data points.}
 \label{fig:NP_plots}
\end{figure}

Normal probability plots of the three normalized residuals are shown in Fig. \ref{fig:NP_plots}. For detailed explanations of properties of such plots, see Filliben and Heckert \cite{Filliben_2013}. Briefly, the residual values $R_i$ are sorted in increasing order and numbered sequentially with their index $i$. Then they are plotted against the values of the function $G(U_i)$, which is the percent point function of the uniform order statistic median of the normal distribution. If the statistical distribution function of the measurements is close to normal, the points on the plot lie close to a straight line with a slope of 1.0 crossing the origin. All three plots in Fig. \ref{fig:NP_plots} are indeed close to a straight line crossing the origin, but the slopes are significantly less than 1.0. This may indicate that the total uncertainties of the energy differences used in the calculation of these normalized residuals are overestimated by a factor between 1.5 and 2. Those uncertainties are in most cases dominated by the measurement uncertainties. Their overestimation may be due to a significant contribution of systematic uncertainties. The plot of $R_{\text{e}-\text{em}}$ (Fig. \ref{fig:NP_plots}, top left) shows that the measurements themselves are internally consistent, i.e., there are no abnormally large discrepancies between individual measurements and their weighted means. The plot of $R_{\text{em}-\text{t}}$ (Fig. \ref{fig:NP_plots}, top right) shows that the mean experimental energies are all statistically consistent with theory, although the two points in the right upper corner of the plot visibly deviate from the overall linear trend. These two points correspond to the level 1s($^2$S)2s2p($^3$P$^{\circ}$) $^4$P$^{\circ}_{5/2}$ (5o1) in Ar~XVI and 1s($^2$S)2s2p($^3$P$^{\circ}$) $^2$P$^{\circ}_{3/2}$ (3o2) in Co~XXV. They will be discussed further below. 

The third normal probability plot (Fig. \ref{fig:NP_plots}, bottom) shows the distribution of differences of all individual measurements from theory. One can see that there are no individual measurements with abnormally large differences from theory. The two points in the right upper corner of this plot correspond to the same levels (5o1 in Ar~XVI and 3o2 in Co~XXV) that are somewhat off the linear trend of $R_{\text{em}-\text{t}}$ discussed above. These are the measurements of Machado et al. \cite{Machado_2020} for Ar~XIV and Smith et al. \cite{Smith_1995} for Co~XXV. Although the deviations of these measurements from theory are entirely within the statistically allowed range for a set of nearly a thousand measurements, it is worth discussing their possible reasons. 

The 5o1 level of Ar~XVI is directly measured by Machado et al. \cite{Machado_2020} as the energy of the U satellite line (see Table \ref{table:sat_labels} for the definition of satellite line labels). Figure 6 of their paper indicates that this measurement might have been affected (shifted to higher energies) by presence of a small additional peak on the wing of the stronger high-energy peak. Such a small peak was indeed detected and measured by Machado et al. in the spectrum of S~XIV (see their Fig. 9 and its discussion in \cite{Machado_2020}). This peak corresponds to the satellite transition i. However, Machado et al. stated that they could not find statistical evidence for the presence of this peak in the argon spectrum. Figure 6(b) of Ref.~\cite{Machado_2020} gives a hint that this might have been due to a combination of a too low signal-to-noise ratio and an imperfection of the modeled shape of the much stronger peak at a higher energy (which is due to Be-like Ar). 

The 3o2 level in Co~XXV is defined by the wavelength of the satellite line q reported by Smith et al. \cite{Smith_1995} to be at 1.72126(11)~{\AA}. According to the calculations presented in Table I of their paper, this line should have been at least 10 times weaker than the satellite `a' located very close to it at the longer-wavelength side. On the short-wavelength side, also within the line width, there is the strong He-like intercombination line y. Figure 3 of Smith et al. indicates that, even if the blending problems might have been mitigated by the use of different beam energies, the signal-to-noise ratio must have been very low for the weak q line, so the small uncertainty assigned to its wavelength was probably underestimated.

Let us now turn to the discussion of measurements that were penalized by large dark uncertainty values in our statistical analysis (see Section \ref{S:stat}). 

As mentioned in Section \ref{S:2}, the measurement of the q+r satellite line in O~VI by Liao et al. \cite{Liao_2013} was found to strongly deviate from the other 14 measurements available for this line. This measurement is responsible for experimental values of two energy levels included in Table \ref{table:2}: 1s($^2$S)2s2p($^3$P$^{\circ}$) $^2$P$^{\circ}$ $J$ = 1/2 (1o2) and $J$ = 3/2 (3o2) at $Z = 8$. Both statistical models of Rukhin \cite{Rukhin_2019,Rukhin_2019a}, CMLE and CRMLE, unanimously singled out this measurement and assigned to it a large dark uncertainty value of about 1500~cm$^{-1}$, which resulted in the total uncertainty $u_{\text{etot}} = 1600$~cm$^{-1}$. This is more than three times greater than the uncertainty specified by Liao et al., 470~cm$^{-1}$. It corresponds to the wavelength uncertainty of 0.008~{\AA}, while Liao et al. stated it to be $(^{+0.0023}_{-0.0020})$~{\AA}. The line in question is marked as O~VI K$\alpha$ in Fig. 3 of Liao et al. (bottom panel) near 22~{\AA}. From this figure, the full width at half maximum is about 0.032~{\AA} for this line. It is unlikely that an error of 0.008~{\AA} could be due to statistical uncertainties of fitting the line profile. There must have been either an error in the calibration of the wavelength scale or some physical phenomenon peculiar to absorption spectra of the interstellar media studied by Liao et al. We note that the wavelength of this line measured by Yao et al. \cite{Yao_2009} and by Gatuzz et al. \cite{Gatuzz_2013} in similar absorption spectra of interstellar media is also too long compared to the laboratory measurements, as well as the calculations of Yerokhin et al. \cite{Y17}, which suggests that a physical explanation (unknown at present) is more likely than a calibration error. The measurements of Yao et al. and Gatuzz et al. had larger uncertainties (0.004~{\AA} and 0.003~{\AA}, respectively) and thus are statistically compatible with the laboratory measurements. The large dark uncertainty assigned to the measurement of Liao et al. makes its contribution to the determination of the experimental mean very small and masks the problem in comparison plots. However, we stress that this is an unresolved problem requiring additional experimental and theoretical study.

Another important case is the measurement of the same q and r satellite transitions in S~XIV by Machado et al. \cite{Machado_2020}. Unlike the oxygen measurements discussed above, these satellite transitions were well resolved in the sulfur spectra observed by Machado et al. Both the CMLE and CRMLE methods assigned relatively large dark uncertainties to these measurements, 390~cm$^{-1}$ for the 1o2 level (r line) and 440~cm$^{-1}$ for the 3o2 level (q line), to be compared with the reported uncertainties of 89~cm$^{-1}$ and 81~cm$^{-1}$, respectively \cite{Machado_2020}. For each of these two levels, there are six measurements included in Table \ref{table:2}. However, only three of them have uncertainties small enough to influence the experimental means. These are the measurements of Machado et al. \cite{Machado_2020}, Schlesser et al. \cite{Schlesser_2013}, and Hell et al. \cite{Hell_2016}. Although in the latter work the q and r lines were unresolved, we extracted the separate values for their upper levels, 3o2 and 1o2, by using their theoretical separation \cite{Y17}. The resulting values agree very well with those reported by Schlesser et al., which leads to singling out the measurements of Machado et al. as discrepant. For these two O~VI levels, the uncertainties of experimental mean energies are a few times smaller than those of the theoretical data \cite{Y17}. Thus, these values are included in the list of the few experimental values that we recommend as reference data. Nevertheless, the discrepancy discussed above calls for an additional experimental study.

All other cases of a non-zero dark uncertainty involve relatively low-resolution measurements in laser-produced, vacuum-spark, tokamak, or EBIT plasmas in $Z= 10$ \cite{Pospieszczyk_1975,Wargelin_2001}, $Z = 21$ \cite{Boiko_1978,Rice_1995,TFR_Group_1985a}, $Z=24$ \cite{TFR_Group_1985a}, $Z = 26$ \cite{Decaux_2003}, and $Z = 28$ \cite{Safronova_1977,Bombarda_1988}. For each of the levels involved, there are only two or three measurements available, and their uncertainties are comparable. The additional dark uncertainties assigned to these measurements range from 0.02 to 1.6 of the $u_{\text{exp}}$ values of Table \ref{table:2}. They may be explained by difficulties in interpretation of partially blended line profiles in these experiments.

\begin{figure}[ht!]
 \centering
 \includegraphics[width=.49\linewidth]{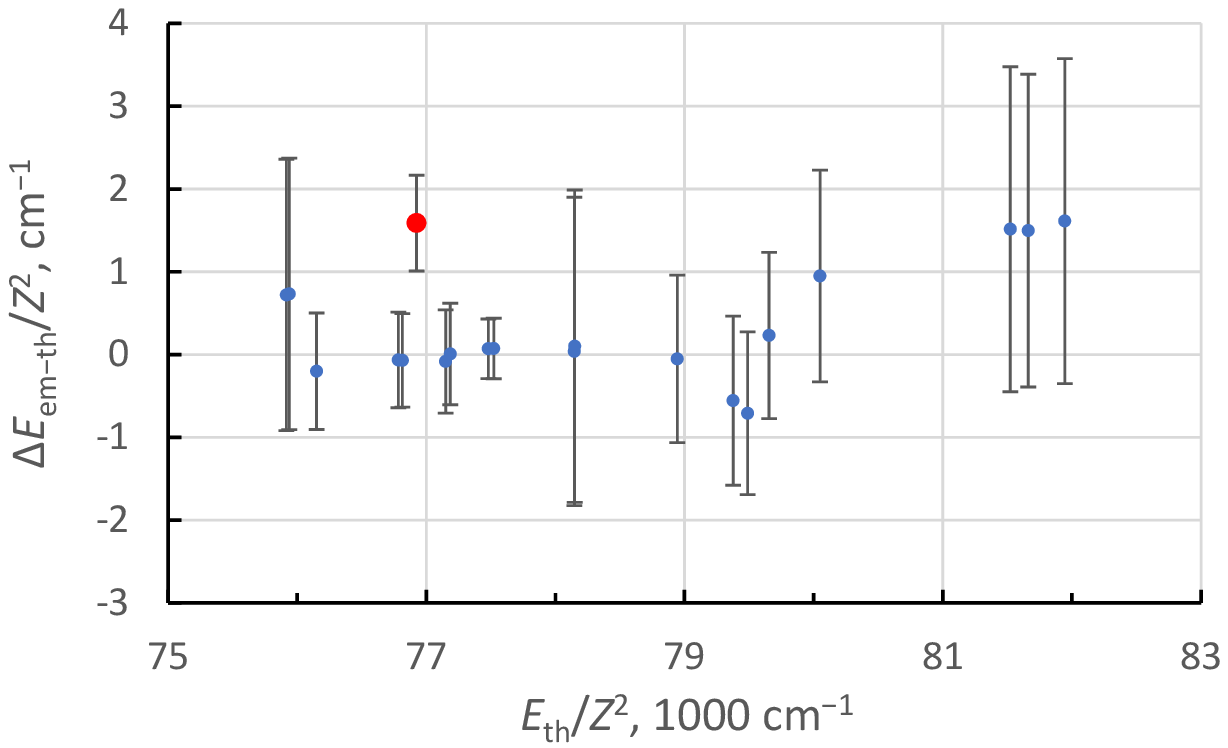}
 \includegraphics[width=.49\linewidth]{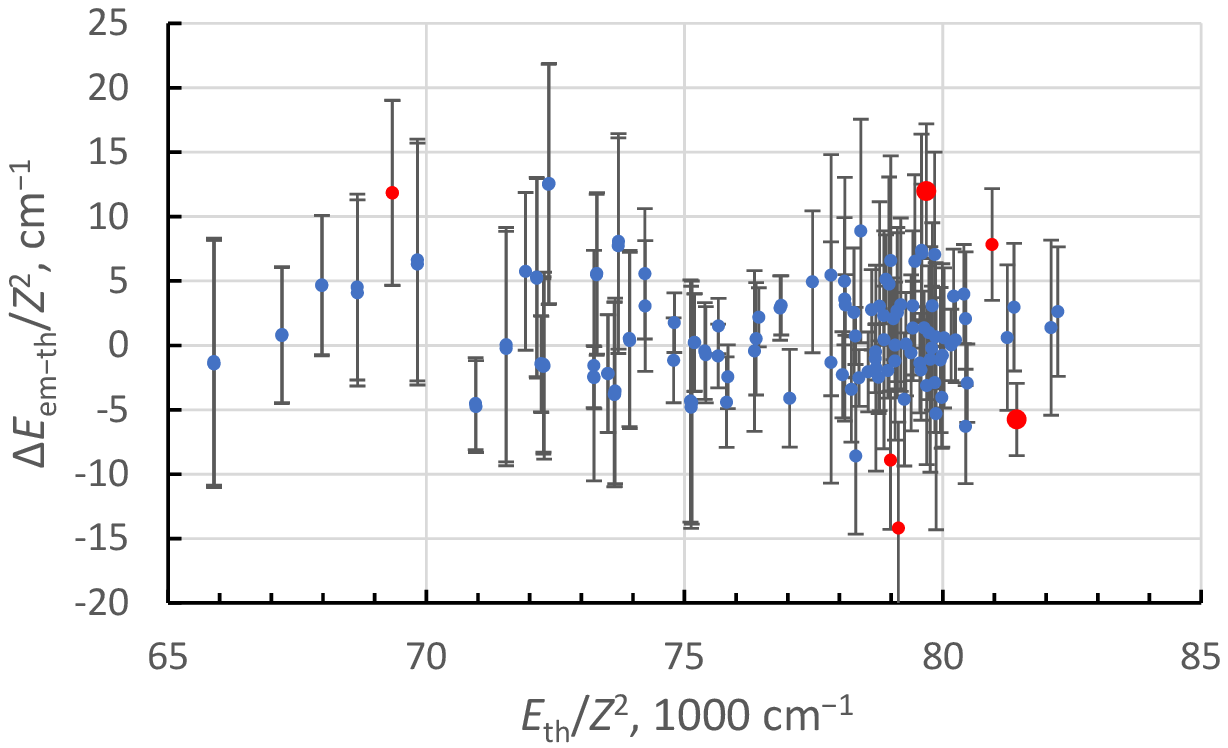}
 \includegraphics[width=.49\linewidth]{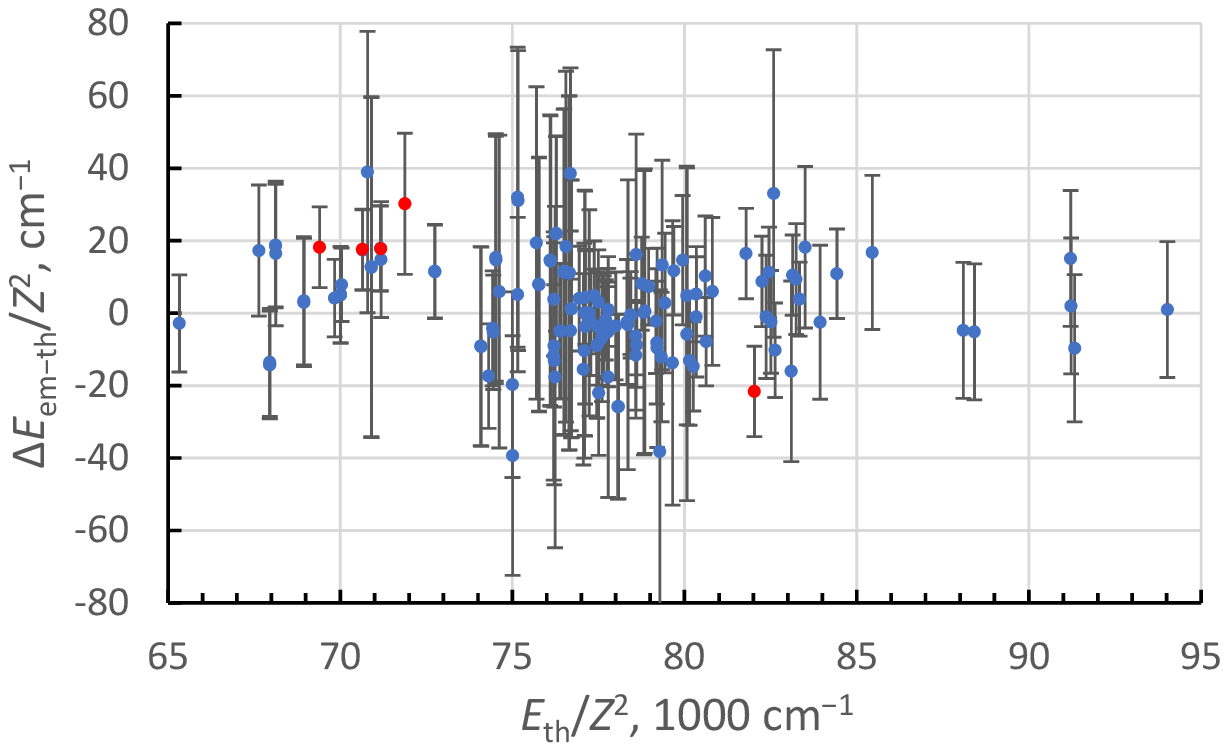}
 \includegraphics[width=.49\linewidth]{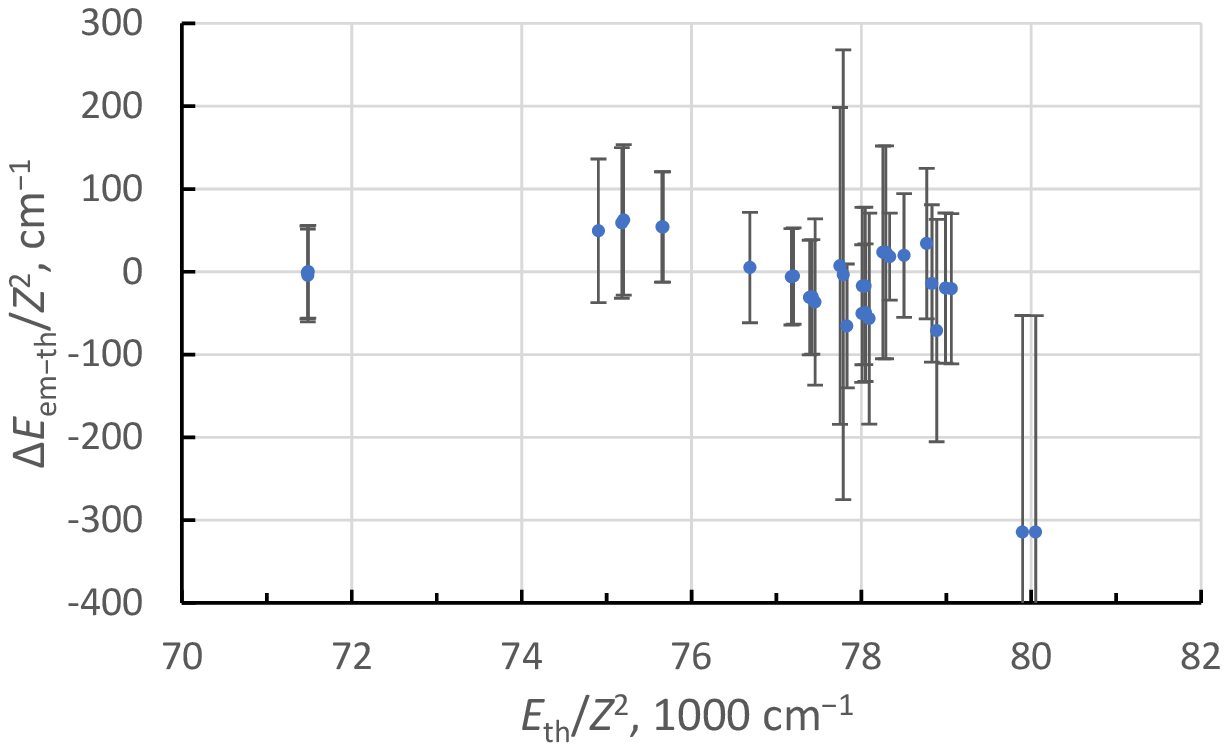}
 \caption{Differences between scaled experimental and theoretical core-excited $n=2$ energy levels of Li-like ions. Experimental values are the mean values ($E_{\text{em}}$) given in Table \ref{table:2}. Theoretical data are from Yerokhin et al. \cite{Y17} for $Z \leq 17$ and from Yerokhin and Surzhykov \cite{Y18} for higher $Z$. The small red dots correspond to the cases of a moderate disagreement ($1 < |R_{\text{em}-\text{t}}| < 1.5$ in Table \ref{table:2}). The large red circles indicate a large disagreement ($|R_{\text{em}-\text{t}}| > 1.5$).}
 \label{fig:dEm-th}
\end{figure}

In addition to the normal probability plots shown in Fig. \ref{fig:NP_plots}, it would be interesting to see a direct comparison of the experimental and theoretical values similar to the plots presented in the review of H- and He-like spectra by Indelicato \cite{Indelicato19}. This turned out difficult because of the large number of measurements (999) and the wide range of their precision (between 1 part in 10$^6$ and 6 parts in 10$^3$). To make plots easier to view, we scaled the measured energies by $Z^{-2}$ and divided all measurements into four sets based on their scaled uncertainties. Instead of making plots for each energy level, we chose to plot data for all levels together. These plots are presented in Fig. \ref{fig:dEm-th}.

The most meaningful comparison of theory with experiment is given by the top left panel of Fig. \ref{fig:dEm-th}, which contains the energy differences ${\Delta}E_{\text{em}-\text{th}}$ with the smallest scaled uncertainties, $u_{\text{sc}} = (u_{\text{exp}}^2 + u_{\text{th}}^2)^{1/2}/Z^2 < 2$~cm$^{-1}$. This panel contains measurements in $Z = 14, 16, 17, 18, 26$, and 36. The ratios of the experimental and theoretical uncertainties, $u_{\text{em}}/u_{\text{th}}$, range between 0.14 and 5.3 for these measurements (see Table \ref{table:2}). In this panel, the only point showing a large discrepancy between experiment and theory corresponds to the measurements of the U transition in Ar~XVI. The mean experimental value for the upper level of this transition, 5o1, is dominated by the high-precision measurement of Machado et al. \cite{Machado_2020}. The possible reasons of its discrepancy with theory have already been discussed above. 

The top right panel of Fig. \ref{fig:dEm-th} represents less precise comparisons with the total uncertainties $u_{\text{sc}}$ in the range between 2~cm$^{-1}$ and 10~cm$^{-1}$. One can see that there are only two strongly discrepant data points in this panel. One of them corresponds to the measurement of the q satellite in $Z=27$ by Smith et al. \cite{Smith_1995}, which was already discussed above. The other one is for the 1s($^2$S)2s2p($^1$P$^{\circ}$) $^2$P$^{\circ}_{3/2}$ (3o3) level in Kr~XXXIV ($Z=36$) by Widmann et al. \cite{Widmann_1995}. This level is defined by the measured wavelength of the `s' satellite, 0.94761(3)~{\AA}. As explained in Section \ref{S:2}, similar to several other studies published by the LLNL group, we adjusted the wavelength calibration of Widmann et al. by subtracting the weighted average difference of the reported He-like lines w, x, y, and z from the precise reference values of Artemyev et al. \cite{Artemyev_2005}. This correction amounted to $-0.000007(11)$~{\AA} and decreased the s-satellite wavelength to 0.94760(3)~{\AA}. This is the value we used to derive the experimental energy of the 3o3 level in Table \ref{table:2}. It is the only measurement available for this level in Kr~XXXIV. Although the correction decreased the discrepancy with Yerokhin and Surzhykov \cite{Y18} from $-2.5\sigma$ to $-2.1\sigma$, it is still significant and calls for additional experimental studies. We note that the theoretical uncertainty for this level \cite{Y18} is seven times smaller than that of Widmann et al., so it is unlikely that the reason for the discrepancy is related to theory. It might be due to blending with some other line, e.g., from a $n = 3$ Li-like satellite transition, which was not suspected in the study of Widmann et al.

The bottom panels of Fig. \ref{fig:dEm-th} show the scaled energy differences between experiment and theory for the much cruder measurements with scaled uncertainties $u_{\text{sc}}$ in the ranges (10--50)~cm$^{-1}$ (left bottom) and (50--270)~cm$^{-1}$ (right bottom). 

Overall, the comparisons shown in Fig. \ref{fig:dEm-th} show a good agreement of experiments with theory. There are only a few measurements that deviate from theory by more than 1.5$\sigma$ (the red circles in the plots), and they are all well within the range of deviations that can be expected for a normal statistical distribution. This corroborates the conclusion following from the normal probability plots shown in Fig. \ref{fig:NP_plots}: the theory of Yerokhin et al. \cite{Y17,Y18} largely agrees with the available experimental data. Thus, we recommend the use of these theoretical data in the cases where more precise experimental data are not available.

However, we must note that the theory \cite{Y17,Y18} is really contested by absolute energy measurement only for $Z = 16$, 17, and 18, where the ratios of experimental and theoretical uncertainties, $u_{\text{em}}/u_{\text{th}}$, are smaller than $\approx 1.5$ (see Table \ref{table:2}). For measurements of the energy intervals within the quartet system, this theory is contested (and confirmed) by experiments for $Z=6$--13 (see subsection \ref{S:3.1}. Therefore, high-precision measurements are still needed to meaningfully test the theory for heavier elements ($Z > 18$), and our recommendation for these elements can only be treated as provisional. 

\section{Conclusion}
\label{S:4}
For the 16 energy levels of the 1s2$l$2$l^{\prime}$ core-excited configurations of Li-like ions, 999 absolute and 35 relative experimental energy-level measurements from 101 publications have been collected and analyzed. Except for a small number of measurements discussed in the previous Sections, agreement between the experimental and theoretical data is very good: all the data obey the normal statistical distribution for the combined (experimental and theoretical) uncertainties. For 35 energy intervals of quartet levels relative to the second lowest quartet level, we provide recommended values derived from experimental data. We also provide recommended absolute experimental energies for seven levels: 1s($^2$S)2s2p($^3$P$^{\circ}$) $^4$P$^{\circ}_{5/2}$ (5o1) in S~XIV, as well as the levels 1s($^2$S)2s2p($^3$P$^{\circ}$) $^2$P$^{\circ}_{1/2,3/2}$ (1o2 and 3o2) in S~XIV, Cl~XV, and Ar~XVI. For the rest of the 1s2$l$2$l^{\prime}$ levels in all Li-like spectra of the elements between carbon and uranium, we recommend the use of the theoretical data of Yerokhin et al. \cite{Y17,Y18} for modeling plasma spectra and for calibration of experimental X-ray spectral measurements in ions with a larger number of electrons. The validity of these theoretical data is meaningfully tested and confirmed for nuclear charges $Z \leq 18$. For higher $Z$, although all available measurements also agree with this theory, their precision is insufficient for a meaningful comparison. High-precision measurements are still needed to contest the theory in this high-$Z$ region.

Several problems in interpretation of experimental data have been detected. Some of them have been solved by re-calibration of experimental wavelength scale and correcting the identification of observed spectral features. However, a few cases require additional experimental and/or theoretical work. One example is the absorption spectrum of Li-like oxygen observed in interstellar media studied with the Chandra orbital X-ray laboratory \cite{Yao_2009,Gatuzz_2013,Liao_2013}. Another example is the absorption spectrum measurements using synchrotrons \cite{Rudolph_2013,Al_Shorman_2013,McLaughlin_2017}. Such experiments can potentially reach very high precision, but this requires an improvement in the instrumentation used.

A new framework of statistical treatment of experimental data has been developed and successfully tested in this work. The newly designed statistics module for Excel implements the recently developed statistical methods for estimation of hidden errors in experiments by using the clustered maximum likelihood and restricted maximum likelihood estimators. This implementation was found to work efficiently in comparisons of up to 22 independent measurements of a single quantity. For a greater number of measurements, the number of possible combinations that needs to be tested to find the optimal division of the data pool into clusters becomes too large to be tractable by these methods. More research is needed to further develop methods for statistical treatment of large sets of measurements.

\ack
The work of one of us (V.I.A) was supported by the contract 1333ND18DNB630011 from the National Institute of Standards and Technology, Gaithersburg, MD 20899, USA. We gratefully acknowledge the help given us by Dr. Andrew L. Rukhin and Gregory W. Haber of the NIST Statistical Engineering Division in implementing computer codes for clustered maximum likelihood uncertainty estimation. We also thank Dr. Beiersforfer of LLNL, Drs. D. Cubaynes and J.-M. Bizau of Universit\'e Paris-Saclay, France, Dr. J. R. Crespo L\'opez-Urrutia of Max-Planck-Institut für Kernphysik, Heidelberg, Germany, and Dr. M. A. Leutenegger of NASA/Goddard Space Flight Center, Greenbelt, USA for helpful communications.




\bibliographystyle{elsarticle-num}
\bibliography{Li-like_sat.bib}







\clearpage
\newpage
\TableExplanation
\section*{Table \ref{table:quart_E}.\label{tbl_qe_te} Recommended theoretical and experimental energies of core-excited 1s2$l$2$l^{\prime}$ quartet levels in Li-like ions.\\
Experimental energies are derived from beam-foil measurements of separations of levels relative to the 1s2s2p~$^4$P$^{\circ}_{3/2}$ level.}
\begin{tabular}{ll} 
$Z$ & Atomic number\\
Lbl & Level label in Table \ref{table:labels}\\
Configuration & Configuration label in $LS$ coupling\\
Term & Term label in $LS$ coupling\\
$J$ & Total angular momentum quantum number\\
$E_{\text{th}}$ & Calculated energy (cm$^{-1}$) \cite{Y17}\\
$u_{\text{th}}$ & Uncertainty of the calculated energy (cm$^{-1}$) \cite{Y17}\\
$E_{\text{exp}}$ & Experimental energy (cm$^{-1}$)\\
$u_{\text{rel}}$ & Uncertainty of the experimental separation from the 1s2s2p $^4$P$^{\circ}_{3/2}$ base level (cm$^{-1}$).\\
 & To obtain the total uncertainty of excitation energy from the ground level, add $u_{\text{th}}$ of the\\
 & base level in quadrature. \\
References & Reference to studies used to derive the level value. TW means `This Work'. \\
\end{tabular} 

\section*{Table \ref{table:quart_wl}.\label{tbl_qw_te} 
Experimental and recommended (Ritz) wavenumbers for transitions between core-excited 1s2$l$2$l^{\prime}$ quartet levels in Li-like ions.\\
Observed wavelengths and wavenumbers are derived as weighted means of beam-foil measurements.\\
Recommended Ritz wavenumbers are derived in a least-squares level optimization in this work.}
\begin{tabular}{ll} 
$Z$ & Atomic number\\
Lbl1 & Lower level label in Table \ref{table:labels}\\
Lbl2 & Upper level label in Table \ref{table:labels}\\
$\lambda_{\text{obs}}$ & Observed wavelength in {\AA}\\
$u(\lambda_{\text{obs}})$ & Uncertainty of observed wavelength in {\AA}\\
$\sigma_{\text{obs}}$ & Observed wavenumber in cm$^{-1}$\\
$u(\sigma_{\text{obs}})$ & Uncertainty of observed wavenumber in cm$^{-1}$\\
$\sigma_{\text{Ritz}}$ & Ritz wavenumber in cm$^{-1}$. This is the recommended experimental value derived in this \\
 & work.\\
$u(\sigma_{\text{Ritz}})$ & Uncertainty of Ritz wavenumber in cm$^{-1}$\\
$E_1$ & Lower level value in cm$^{-1}$ from Table \ref{table:quart_E}\\
$E_2$ & Upper level value in cm$^{-1}$ from Table \ref{table:quart_E}\\
F & Line flag: P -- predicted wavenumber; b -- blended line\\
Weight & Transition weight in the level optimization procedure (1 for unblended lines, 0 for lines \\
 & excluded from the level optimization, other values for blended lines)\\
References & References to line observations used to derive the weighted mean $\lambda_{\text{obs}}$\\
\end{tabular} 

\section*{Table \ref{table:2}.\label{tbl2te} Comparison of theoretical and experimental energies of the core-excited 1s2$l$2$l^{\prime}$ states in Li-like ions.}
\begin{tabular}{ll} 
$Z$ & Atomic number\\
$E_{\text{th}}$ & Calculated energy (cm$^{-1}$) \cite{Y17,Y18}\\
$U_{\text{th}}$ & Uncertainty of the calculated energy (cm$^{-1}$) \cite{Y17,Y18}\\
$E_{\text{exp}}$ & Experimental energy (cm$^{-1}$)\\
$U_{\text{exp}}$ & Uncertainty of the experimental energy (cm$^{-1}$) \\
${\Delta}$ & Difference between the experimental and calculated energy values, \\
& $E_{\text{exp}} - E_{\text{th}}$ (cm$^{-1}$) \\
$R$ & Ratio ${\Delta}$/$U_{\text{exp}}$ of the difference between the experimental \\
& and calculated energy values to the uncertainty of the experimental energy \\
$E_{\text{exp,mean}}$ & Mean experimental energy (cm$^{-1}$), averaged over all values $E_{\text{exp}}$ \\
& with the same $Z$ using weights equal to $U_{\text{exp}}^{-2}$ \\
$U_{\text{exp,mean}}$ & Uncertainty of the mean experimental energy (cm$^{-1}$) \\
${\Delta}_{\text{mean}}$ & Difference between the mean experimental and calculated energy values, \\
& $E_{\text{exp,mean}} - E_{\text{th}}$ (cm$^{-1}$) \\
$U_{\text{mean}}$ & Uncertainty of the difference between the mean experimental and \\
& calculated energy values (cm$^{-1}$), with $U_{\text{th}}$ included \\
$R_{\text{mean}}$ & Ratio ${\Delta}_{\text{mean}}$/$U_{\text{mean}}$ of the difference between the mean experimental \\
& and calculated energy values to the uncertainty of the difference \\
$N$ & The number of the data point (measurement number) for the given level \\
Ref & Reference \\
\end{tabular} 

\datatables 
\LTright=0pt
\LTleft=0pt

\footnotesize

\begin{center}
\begin{longtable}{@{\extracolsep\fill}cccccccccc}
  \caption{Recommended theoretical and experimental energies of core-excited 1s2$l$2$l^{\prime}$ quartet levels in Li-like ions.\\
  Experimental energies are derived from beam-foil measurements of separations of levels relative to the 1s2s2p~$^4$P$^{\circ}_{3/2}$ level.\\ 
  See page \pageref{tbl_qe_te} for Explanation of Tables.\label{table:quart_E}}
\multirow[m]{2}{*}{$Z$} & \multirow[m]{2}{*}{Lbl} & \multirow[m]{2}{*}{Configuration} & \multirow[m]{2}{*}{Term} & \multirow[m]{2}{*}{$J$} & $E_{\text{th}}$ & $u_{\text{th}}$ & $E_{\text{exp}}$ & $u_{\text{rel}}$ & \multirow[m]{2}{*}{References}\\
 & & & & & cm$^{-1}$ & cm$^{-1}$ & cm$^{-1}$ & cm$^{-1}$ & \\
\hline\noalign{\vskip3pt}
\endfirsthead
\caption[]{(continued)}
\multirow[m]{2}{*}{$Z$} & \multirow[m]{2}{*}{Lbl} & \multirow[m]{2}{*}{Configuration} & \multirow[m]{2}{*}{Term} & \multirow[m]{2}{*}{$J$} & $E_{\text{th}}$ & $u_{\text{th}}$ & $E_{\text{exp}}$ & $u_{\text{rel}}$ & \multirow[m]{2}{*}{Ref}\\
 & & & & & cm$^{-1}$ & cm$^{-1}$ & cm$^{-1}$ & cm$^{-1}$ & \\
\hline\noalign{\vskip3pt}
\endhead
\hline\noalign{\vskip3pt}
\multicolumn{10}{r}{\itshape (continued on next page)}\\
\endfoot
\noalign{\vskip3pt}\hline
\endlastfoot
6  & 1o1 & 1s($^2$S)2s2p($^3$P$^{\circ}$) & $^4$P$^{\circ}$ & 1/2 & 2371985  & 9  & 2371987.9 & 2.8 & \cite{Livingston_1978,Engstrom_1987,Y17},TW   \\
6  & 3o1 & 1s($^2$S)2s2p($^3$P$^{\circ}$) & $^4$P$^{\circ}$ & 3/2 & 2371989  & 9  & 2371990.6 & 0.0 & \cite{Livingston_1978,Engstrom_1987,Y17},TW   \\
6  & 5o1 & 1s($^2$S)2s2p($^3$P$^{\circ}$) & $^4$P$^{\circ}$ & 5/2 & 2372086  & 9  & 2372086.8 & 2.3 & \cite{Livingston_1978,Engstrom_1987,Y17},TW   \\
6  & 1e2 & 1s($^2$S)2p$^2$($^3$P)         & $^4$P           & 1/2 & 2446336  & 9  & 2446332.3 & 2.1 & \cite{Livingston_1978,Engstrom_1987,Y17},TW   \\
6  & 3e1 & 1s($^2$S)2p$^2$($^3$P)         & $^4$P           & 3/2 & 2446412  & 9  & 2446409.4 & 2.4 & \cite{Livingston_1978,Engstrom_1987,Y17},TW   \\
6  & 5e1 & 1s($^2$S)2p$^2$($^3$P)         & $^4$P           & 5/2 & 2446451  & 9  & 2446452.3 & 1.8 & \cite{Livingston_1978,Engstrom_1987,Y17},TW   \\
7  & 1o1 & 1s($^2$S)2s2p($^3$P$^{\circ}$) & $^4$P$^{\circ}$ & 1/2 & 3337990  & 11 & 3337980   & 8   & \cite{Livingston_1978,Y17},TW                 \\
7  & 3o1 & 1s($^2$S)2s2p($^3$P$^{\circ}$) & $^4$P$^{\circ}$ & 3/2 & 3338024  & 11 & 3338015   & 0   & \cite{Livingston_1978,Y17},TW                 \\
\end{longtable} 
\end{center}

\begin{center}
\begin{longtable}{@{\extracolsep\fill}ccccccccccccccc}
  \caption{Experimental and recommended (Ritz) wavenumbers for transitions between core-excited 1s2$l$2$l^{\prime}$ quartet levels in Li-like ions.\\
  Observed wavelengths and wavenumbers are derived as weighted means of beam-foil measurements.\\
  Recommended Ritz wavenumbers are derived in a least-squares level optimization in this work.  
  See page \pageref{tbl_qw_te} for Explanation of Tables.\label{table:quart_wl}}
\multirow[m]{2}{*}{$Z$} & \multirow[m]{2}{*}{Lbl1} & \multirow[m]{2}{*}{Lbl2} & $\lambda_{\text{obs}}$ & $u(\lambda_{\text{obs}})$ & $\sigma_{\text{obs}}$ & $u(\sigma_{\text{obs}})$ & $\sigma_{\text{Ritz}}$ & $u(\sigma_{\text{Ritz}})$ & $\Delta\sigma_{\text{obs}-\text{Ritz}}$ & $E_1$ & $E_2$ & \multirow[m]{2}{*}{F} & \multirow[m]{2}{*}{Weight} & \multirow[m]{2}{*}{References}\\
 & & & {\AA} & {\AA} & cm$^{-1}$ & cm$^{-1}$ & cm$^{-1}$ & cm$^{-1}$ & cm$^{-1}$ & cm$^{-1}$ & cm$^{-1}$ & & & \\
\hline\noalign{\vskip3pt}
\endfirsthead
\caption[]{(continued)}
\multirow[m]{2}{*}{$Z$} & \multirow[m]{2}{*}{Lbl1} & \multirow[m]{2}{*}{Lbl2} & $\lambda_{\text{obs}}$ & $u(\lambda_{\text{obs}})$ & $\sigma_{\text{obs}}$ & $u(\sigma_{\text{obs}})$ & $\sigma_{\text{Ritz}}$ & $u(\sigma_{\text{Ritz}})$ & $\Delta\sigma_{\text{obs}-\text{Ritz}}$ & $E_1$ & $E_2$ & \multirow[m]{2}{*}{F} & \multirow[m]{2}{*}{Weight} & \multirow[m]{2}{*}{References}\\
 & & & {\AA} & {\AA} & cm$^{-1}$ & cm$^{-1}$ & cm$^{-1}$ & cm$^{-1}$ & cm$^{-1}$ & cm$^{-1}$ & cm$^{-1}$ & & & \\
\hline\noalign{\vskip3pt}
\endhead
\hline\noalign{\vskip3pt}
\multicolumn{15}{r}{\itshape (continued on next page)}\\
\endfoot
\noalign{\vskip3pt}\hline
\endlastfoot
6  & 3o1 & 5e1 & 1342.990 & 0.037 & 74460.7  & 2.1 & 74461.7 & 1.8 &$-1$  & 2371990.6 & 2446452.3 &   & 1     & \cite{Livingston_1978,Engstrom_1987}\\
6  & 1o1 & 3e1 & 1343.690 & 0.037 & 74421.9  & 2.3 & 74421.5 & 2.1 &$0.4$ & 2371987.9 & 2446409.4 &   & 0.781 & \cite{Livingston_1978,Engstrom_1987}\\
6  & 3o1 & 3e1 & 1343.690 & 0.037 & 74421.9  & 4.4 & 74418.8 & 2.4 &$3.1$ & 2371990.6 & 2446409.4 &   & 0.219 & \cite{Livingston_1978,Engstrom_1987}\\
6  & 5o1 & 5e1 & 1344.690 & 0.037 & 74366.6  & 2.0 & 74365.5 & 1.8 &$1.1$ & 2372086.8 & 2446452.3 &   & 1     & \cite{Livingston_1978,Engstrom_1987}\\
6  & 1o1 & 1e2 & 1345.133 & 0.037 & 74342.1  & 4.4 & 74344.4 & 3.0 &$-2.3$& 2371987.9 & 2446332.3 &   & 0.214 & \cite{Livingston_1978,Engstrom_1987}\\
6  & 3o1 & 1e2 & 1345.133 & 0.037 & 74342.1  & 2.3 & 74341.7 & 2.1 &$0.4$ & 2371990.6 & 2446332.3 &   & 0.786 & \cite{Livingston_1978,Engstrom_1987}\\
6  & 5o1 & 3e1 & 1345.507 & 0.037 & 74321.4  & 2.0 & 74322.6 & 1.8 &$-1.2$& 2372086.8 & 2446409.4 &   & 1     & \cite{Livingston_1978,Engstrom_1987}\\
6  & 1e2 & 3e1 &          &       &          &     & 77.1    & 2.8 &      & 2446332.3 & 2446409.4 & P & 0     & \\
\end{longtable} 
\end{center}

\begin{center}
\begin{landscape}
\begin{longtable}{@{\extracolsep\fill}ccccccccccccccc}
  \caption{Comparison of theoretical and experimental energies of core-excited 1s2$l$2$l^{\prime}$ states in Li-like ions. See page\ \pageref{tbl2te} for Explanation of Tables\label{table:2}}
\multirow[m]{2}{*}{$Z$} & \multirow[m]{2}{*}{$N_{\text{m}}$} & $E_{\text{th}}$ & $u_{\text{th}}$ & $E_{\text{exp}}$ & $u_{\text{exp}}$ & $u_{\text{dark}}$ & $u_{\text{etot}}$ & $E_{\text{em}}$ & $u_{\text{em}}$ & \multirow[m]{2}{*}{$u_{\text{em}}/u_{\text{th}}$} & \multirow[m]{2}{*}{$R_{\text{e}-\text{em}}$} & \multirow[m]{2}{*}{$R_{\text{em}-\text{t}}$} & \multirow[m]{2}{*}{$R_{\text{e}-\text{t}}$} & \multirow[m]{2}{*}{Ref}\\
 & & cm$^{-1}$ & cm$^{-1}$ & cm$^{-1}$ & cm$^{-1}$ & cm$^{-1}$ & cm$^{-1}$ & cm$^{-1}$ & cm$^{-1}$ \\
\hline\noalign{\vskip3pt}
\endfirsthead
\caption[]{(continued)}
\multirow[m]{2}{*}{$Z$} & \multirow[m]{2}{*}{$N_{\text{m}}$} & $E_{\text{th}}$ & $u_{\text{th}}$ & $E_{\text{exp}}$ & $u_{\text{exp}}$ & $u_{\text{dark}}$ & $u_{\text{etot}}$ & $E_{\text{em}}$ & $u_{\text{em}}$ & \multirow[m]{2}{*}{$u_{\text{em}}/u_{\text{th}}$} & \multirow[m]{2}{*}{$R_{\text{e}-\text{em}}$} & \multirow[m]{2}{*}{$R_{\text{em}-\text{t}}$} & \multirow[m]{2}{*}{$R_{\text{e}-\text{t}}$} & \multirow[m]{2}{*}{Ref}\\
 & & cm$^{-1}$ & cm$^{-1}$ & cm$^{-1}$ & cm$^{-1}$ & cm$^{-1}$ & cm$^{-1}$ & cm$^{-1}$ & cm$^{-1}$ \\
\hline\noalign{\vskip3pt}
\endhead
\hline\noalign{\vskip3pt}
\multicolumn{14}{r}{\itshape (continued on next page)}\\
\endfoot
\noalign{\vskip3pt}\hline
\endlastfoot
\multicolumn{15}{l}{1e1: 1s2s$^2$ $^2$S$_{1/2}$ or [1s2s$^2$]$_{1/2}$}\\
6  & 1 & 2351890   & 66 & 2355900 & 4000 & 0 & 4000 & 2351790 & 480 & 7.3 & 1.03 & -0.21 & 1.00 & \cite{Rodbro_1979}\\
6  & 2 & 2351890   & 66 & 2355100 & 8100 & 0 & 8100 & 2351790 & 480 &  & 0.41 &  & 0.40 & \cite{Schneider_1977}\\
6  & 3 & 2351890   & 66 & 2351860 & 810 & 0 & 810 & 2351790 & 480 &  & 0.09 &  & -0.04 & \cite{Mann_1987}\\
6  & 4 & 2351890   & 66 & 2350200 & 4400 & 0 & 4400 & 2351790 & 480 &  & -0.36 &  & -0.38 & \cite{Hofmann_1990}\\
6  & 5 & 2351890   & 66 & 2351710 & 860 & 0 & 860 & 2351790 & 480 &  & -0.09 &  & -0.21 & \cite{Mannervik_1997}\\
6  & 6 & 2351890   & 66 & 2351460 & 990 & 0 & 1000 & 2351790 & 480 &  & -0.33 &  & -0.43 & \cite{Mack_1987}\\
6  & 7 & 2351890   & 66 & 2352100 & 1700 & 0 & 1700 & 2351790 & 480 &  & 0.18 &  & 0.12 & \cite{Lee_1992}\\
\end{longtable} 
\end{landscape}
\end{center}

\clearpage

\end{document}